\documentclass[a4paper,fleqn,usenatbib]{mnras}

\usepackage[T1]{fontenc}
\usepackage{ae,aecompl}

\usepackage{graphicx}	% Including figure files
\usepackage{amsmath}	% Advanced maths commands
\usepackage{amssymb}	% Extra maths symbols
\usepackage[usenames,dvipsnames]{color}
\usepackage{siunitx}

\newcommand{\cena}{NGC\,5128}
\newcommand{\btrfrac}{$\xi_{\rm b/r}$}

\newcommand{\tr}[1]{\textcolor{Red}{#1}}

\title[The Extended Globular Cluster System of \cena]{The Survey of Centaurus\,A's Baryonic Structures (SCABS). II. The Extended Globular Cluster System of \cena\ and its Nearby Environment}

\author[M. A. Taylor et al.]{Matthew A. Taylor$^{1,2}$\thanks{E-mail: mtaylor@astro.puc.cl (MAT)},
Thomas H. Puzia$^{1}$,
Roberto P. Mu\~noz$^{1}$,
Steffen Mieske$^{2}$,
\newauthor{Ariane Lan\c{c}on$^{3}$,
Hongxin Zhang$^{1}$,
Paul Eigenthaler$^{1}$,
and Mia Sauda Bovill$^{4}$}
\\
$^{1}$Institute of Astrophysics, Pontificia Universidad Cat\'olica de Chile, Av.~Vicu\~na Mackenna 4860, 7820436 Macul, Santiago, Chile\\
$^{2}$European Southern Observatory, Alonso de Cordova 3107, Vitacura, Santiago, Chile\\
$^{3}$Observatoire astronomique de Strasbourg, Universit\'e de Strasbourg, CNRS, UMR 7550, 11 rue de l'Universite, F-67000 Strasbourg, France\\
$^{4}$Space Telescope Science Institute, 3700 San Martin Drive, 21218, Baltimore, Maryland, USA
}

\date{Accepted XXX. Received YYY; in original form \today}

\pubyear{2016}

\begin{document}
\label{firstpage}
\pagerange{\pageref{firstpage}--\pageref{lastpage}}
\maketitle

% Abstract of the paper
\begin{abstract}
New wide-field $u'g'r'i'z'$ {\it Dark Energy Camera} observations centred on the nearby giant elliptical galaxy \cena\ covering $\sim21\,{\rm deg}^2$ are used to compile a new catalogue of $\sim3\,200$ globular clusters (GCs). We report 2\,404 newly identified candidates, including the vast majority within $\sim140$\,kpc of \cena. We find evidence for a transition at a galactocentric radius of $R_{\rm gc}\approx55$\,kpc from GCs ``intrinsic'' to \cena\ to those likely to have been accreted from dwarf galaxies or that may transition to the intra-group medium of the Centaurus A galaxy group. We fit power-law surface number density profiles of the form $\Sigma_{N, R_{\rm gc}}\propto R_{\rm gc}^\Gamma$ and find that inside the transition radius, the red GCs are more centrally concentrated than the blue, with $\Gamma_{\rm inner,red}\approx-1.78$ and $\Gamma_{\rm inner,blue}\approx-1.40$, respectively. Outside this region both profiles flatten, more dramatically for the red GCs ($\Gamma_{\rm outer,red}\approx-0.33$) compared to the blue ($\Gamma_{\rm outer,blue}\approx-0.61$), although the former is more likely to suffer contamination by background sources. The median $(g'\!-\!z')_0\!=\!1.27$\,mag colour of the inner red population is consistent with arising from the amalgamation of two giant galaxies each less luminous than present-day \cena. Both in- and out-ward of the transition radius, we find the fraction of blue GCs to dominate over the red GCs, indicating a lively history of minor-mergers. Assuming the blue GCs to originate primarily in dwarf galaxies, we model the population required to explain them, while remaining consistent with \cena's present-day spheroid luminosity. We find that that several dozen dwarfs of luminosities $L_{{\rm dw},V}\simeq10^{6-9.3}\,L_{V,\odot}$, following a Schechter luminosity function with a faint-end slope of $-1.50\lesssim\alpha\lesssim-1.25$ is favoured, many of which may have already been disrupted in \cena's tidal field.
\end{abstract}

% Select between one and six entries from the list of approved keywords.
% Don't make up new ones.

%%%%%%%%%%%%%%%%%%%%%%%%%%%%%%%%%%%%%%%%%%%%%%%%%%

%%%%%%%%%%%%%%%%% BODY OF PAPER %%%%%%%%%%%%%%%%%%

\section{Introduction}
\begin{figure*}
	\includegraphics[width=0.9\linewidth]{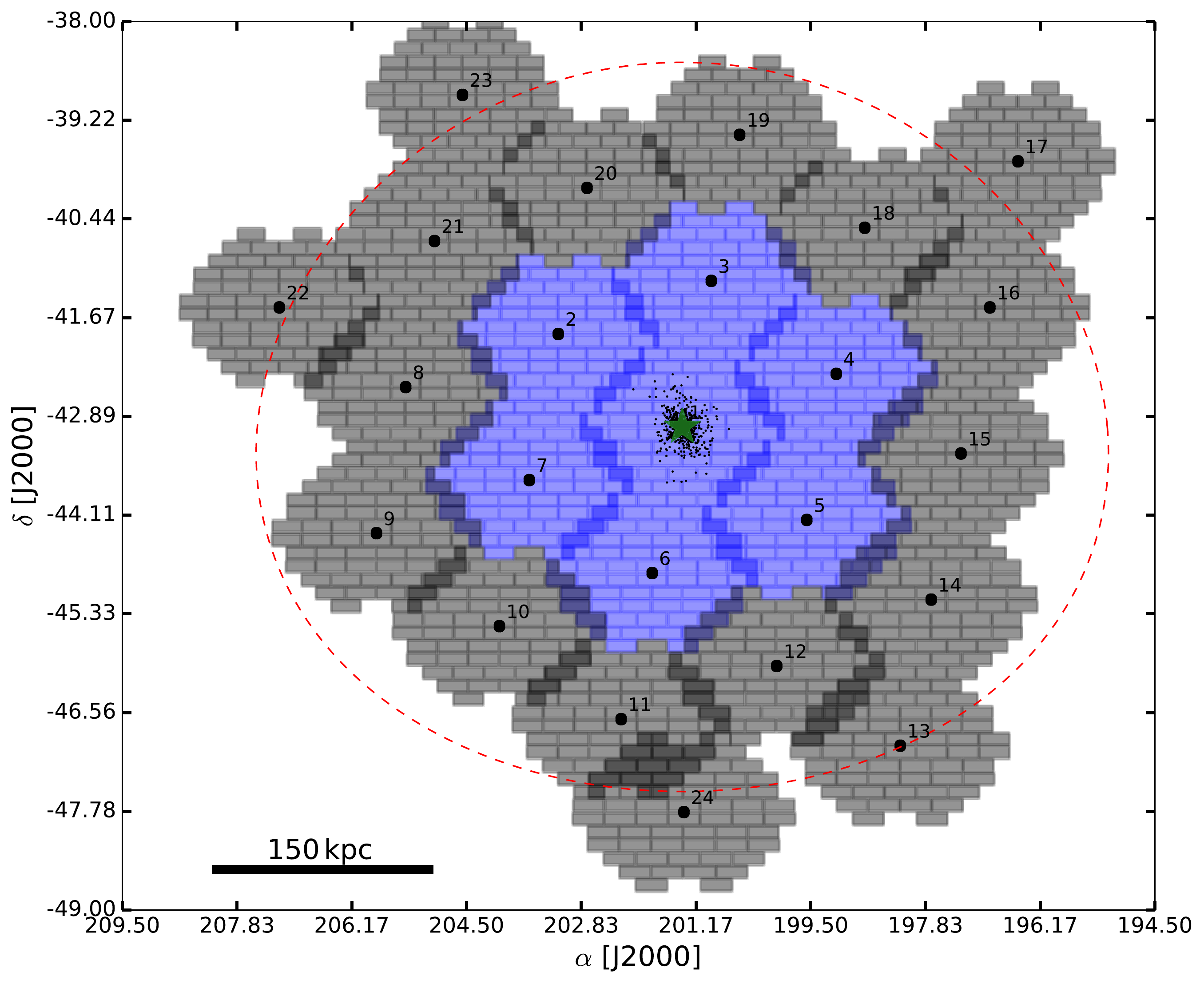}
	\caption{The spatial footprint of the SCABS observations, with the fields studied in this work marked as blue tiles. The position of \cena\ is shown by the green star, while the surrounding cloud of black dots indicates the population of confirmed GCs. The red dashed ellipse shows the $\sim300$\,kpc extent of \cena's virial radius, and the tiles considered in this work cover $\sim120-150$\,kpc in galactocentric radius. Different tiles are indicated by the numbers shown, with tile\,1 centred on \cena\ itself.}
	\label{fig:dither}
\end{figure*} 

Resolved stellar population studies are the best direct probes of a galaxy's evolutionary past, but today's instrumentation limits these techniques to galaxies just outside the Local Group. Luckily, the formation environment of a galaxy is forever encoded in the ubiquitous globular cluster (GC) systems that surround them. Stars generally form in a clustered fashion governed by their environmental conditions \citep[e.g.][]{lad03, mlk04, por10}, which dissolve on Gyr timescales for low-mass clusters (${\cal M}_\star\la10^5$), but the compact configurations of massive clusters (${\cal M}_\star\ga10^5$) enable their survival, while their relatively high luminosities \citep[$M_V\!\lesssim\!-7.5$\,mag;][]{har01} make them observationally accessible in the local universe. GCs are the best approximations to single-aged, chemically homogeneous stellar populations given that they form during periods of intense star formation early in a galaxy's history, or later during merger-induced starbursts. These features make GCs very useful probes of the galactic conditions in which they form \citep[e.g.][]{har91,ash98,bro06,ash08,pen08,geo10}.

The well-known bimodal colour/metallicity distributions of GCs \cite[e.g.][]{sea78,zep93,ost93,whi95,els96,geb99,puz99,puz04,puz05,puz06,kun01,lar01,pen06,pen11,spi06,gou07, brod12, ush12, ush15} act as probes of galactic formation histories \cite[e.g.][for a review]{bro06}. This bimodality has been interpreted in two ways. The merger scenario sees the blue/metal-poor GCs representing those that formed in-situ from pristine gas along with their giant hosts, while the red/metal-rich peak corresponds to GCs that formed later in gas-rich merger induced starbursts \citep[e.g.][]{ash92,kis97,for97b,bea02,spi06}. Conversely, in the hierarchical build-up scenario, it may be that the red GCs are the primordial GCs that form and are rapidly enriched in the dense star-forming environments of their giant hosts, while blue GCs are accreted from dwarf galaxies during minor-merger events \citep[e.g.][]{cot98,cot00,dab14a,dab14b,dab15}. The latter interpretation seems to be supported by the observation that red GCs often coincide with the underlying bulge light of their giant hosts, while the blue GCs tend to be more prevalent in the outer galaxy halos \citep[e.g.][]{gei96,for97b,ash98,cot98,cot00,for01,bas06,bro06,spi06,gou07,fai11,for12,dab14a,dab14b,dab15,kar16}, as well as the fact that dwarf galaxies are known to host primarily metal-poor GCs \citep{pen06, puz08} that would tend to deposit their GCs in the giants host's halo. In any case, knowledge of the distributions in both colour and space of both GCs and dwarf galaxies around a giant galaxy provides useful leverage on understanding the ancient environment that gave rise to it.

\cena\ is the dominant giant elliptical (gE) galaxy in the nearby \citep[$3.8\pm0.1\,{\rm Mpc}$;][]{har10a} Centaurus A galaxy group, which is mainly comprised of at least 40 known dwarf galaxies \citep[][]{cot97,van00,kar07,crn14,crn15,tul15}. The task of identifying GCs around \cena\ is non-trivial and the proximity of \cena\ acts as both an advantage and disadvantage. On one hand, even the faintest GCs are observationally accessible; however, \cena's halo spreads across a large portion of the sky, requiring very wide-field imagers to probe into the extreme halo. For this reason, the identification of the GC system has come in fits and starts, with the first confirmed GC identified by \citet{gra80}, and handfuls of GCs found throughout the balance of the 1980s and into the 1990s \citep[e.g.][]{van81,har84a,har84b,hes84,hes86,sha88,har88,min96,hol99}. With the turn of the millennium, GCs finally began to be discovered {\it en masse} with dozens identified through deep imaging with multiple broad-band optical filters \citep[e.g.][]{rej01,pen03,har04a,pen04a,gom06}. More recently, \cite{har12} (hereafter H12) made use of exceptional ($\sim0.5''$) seeing to image 1.55 deg$^2$ ($\sim90\times90\,{\rm kpc}^2$) centred on the galaxy and identified $\gtrsim800$ GC candidates based on $B$- and $R$-band optical photometry. These candidates are of tremendous use for large-scale spectroscopic follow-up, but the uncertain level of foreground and background contamination and restricted spectral energy distribution (SED) coverage limits its utility to characterize the overall GC system of \cena\ with a large degree of confidence.

Regardless, much has already been learned about \cena\ from the $\sim600$ confirmed GCs \citep[][]{van81,hes84,hes86,har92,jab96,hel97,pen04c,woo05,rej07,bea08,woo10a}. For example, in accords with other giant galaxies, \cena's GCs show a bi- and possibly tri-modal colour/metallicity distribution \citep{min96,hel97,har02b,pen04c,bea08,spi08,sin10,woo10b}, corresponding to at least two distinct GC populations. There is significant evidence supporting a recent major merger \citep{baa54,gra79,inn79,tub80,mal83,hes86,qui93,min96,sti04}, which may have given rise to at least one new generation of GCs being produced since the earliest years of \cena's past \citep[e.g.][]{van81,hes84,hes86,har92,pen04c,woo05,rej07,bea08,woo10b}. Despite the large amount of knowledge gained from the known GCs around \cena, the seemingly simple question of the total population of GCs is still somewhat uncertain, with total estimates ranging from $\sim\!1\,000\!-\!2\,000$, including as many as $1\,500$ possibly residing within the inner $25'$\ \citep{har84a,har02b,har04b,har06,har10b}. In fact, the spatial distribution shows significant variation \citep{woo07,har12}, indicating that the total population extends beyond $45'$, possibly along a certain preferential axis which may trace unknown tidal features from past mergers.

In this work, we use new CTIO/DECam wide-field, five-band ($u'g'r'i'z'$) photometry to identify GC candidates within $\sim140\,{\rm kpc}$ of \cena. The data cover $\sim21\,{\rm deg}^2$, and are a subset of the $\sim72\,{\rm deg}^2$ ``{\it Survey of Centaurus A's Baryonic Structures}'' (SCABS) imaging campaign presented in the first paper of this series \citep[][hereafter Paper\,I.]{tay16}. In addition to presenting a new list of likely candidates, we also subject the candidate list of H12 to the same selection criteria, showing that many are likely to be stellar in nature. We show via a preliminary analysis of the overall GC system characteristics that our $u'g'r'i'z'$ photometry is sufficient to detect the majority of GC candidates within $\sim140\,{\rm kpc}$ with a high degree of confidence. With these candidates, recent near-infrared (NIR) wide-field imaging and extra spectroscopic confirmations in the near future, a complete census of \cena's GC system is nearly at hand, along with the myriad lines of scientific inquiry that will come with it.

The paper is organized as follows. \S\,\ref{sec:data} gives a brief summary of the SCABS observations upon which this work is based. In \S\ref{sec:analysis} we introduce our novel GC selection technique based on optical colours and source morphologies, including the use of populations of GCs and foreground stars modelled from radial-velocity confirmed samples. \S\,\ref{sec:discussion} discusses the distributions of GCs in colour/space reaching into the extreme halo of \cena. We posit that our GC candidates begin to transition from the GCs ``intrinsic'' to \cena, to the population likely to be associated with the intra-group medium of Centaurus A, and briefly investigate the dwarf galaxy reservoirs, past and present, required to give rise to the population of blue GCs observed. Finally, we briefly conclude and summarize the work in \S\,\ref{sec:conc}. Throughout this work we adopt a distance modulus for \cena\ of $(m\!-\!M)\!=\!27.88\pm0.05\,{\rm mag}$, corresponding to a distance of $3.8\pm0.1\,{\rm Mpc}$ \citep{har10a, har10b}.

%%%%%%%%%%%%%%%%%%%%%%%%%%%%%%%
%%%%%%%%%%%%%%%%%%%%%%%%%%%%%%%
%%%%%%%%%%%%%%%%%%%%%%%%%%%%%%%
\section{The Data}
\label{sec:data}
The analysis presented in this contribution, as well as the results and discussion based on it, are formed upon the SCABS dataset presented and described in Paper\,I. We begin with the catalogue of sources with near-ultraviolet (NUV) and optical $u'g'r'i'z'$ photometry and morphological properties produced by the SCABS data reduction procedures. These data consist of $>500\,000$ sources with complete sets of $u'g'r'i'z'$ photometry, all with on-sky coordinates corresponding to the inner $\sim21\,{\rm deg}^2$ of the Centaurus A galaxy group, reaching out to $\sim150$\,kpc in galacto-centric radius from \cena\ (see Fig.\,\ref{fig:dither}). We note here that, while Tables\,\ref{tbl:point_cat} and \ref{tbl:gccand_cat} list photometric data that are uncorrected for foreground extinction, in practice we de-redden all sources on a point-by-point basis using the extinction maps of \cite{sch11} and use these values for the analysis.

Altogether, this rich dataset is exploited to identify point-like and marginally resolved sources, the numbers of which are reduced by two orders of magnitude to reveal a large catalogue of likely GCs as described in the following.

%%%%%%%%%%%%%%%%%%%%%%%%%%%%%%%%
%%%%%%%%%%%%%%%%%%%%%%%%%%%%%%%%
%%%%%%%%%%%%%%%%%%%%%%%%%%%%%%%%

\section{Analysis}
\label{sec:analysis}
This section describes the process of converting the calibrated SCABS images into catalogues of point-like and marginally resolved sources from which a GC candidate list is selected. The distance to \cena\ of 3.8\,Mpc corresponds to a physical scale of 18\,{\rm pc}/$''$. Given GC sizes of $1-20$\,pc ($0.06-1.11\arcsec$) with a typical average half-light radius of $\sim3$\,pc ($\sim0.12\arcsec$), and typical PSF FWHM of $1-1.5\arcsec$ in all bands during our observations, we expect most GCs to be point-like or marginally resolved with sizes accounting for $4-111$ percent of their PSF FWHM, or $\sim10$ percent for an average GC at $1.25\arcsec$ FWHM.

\begin{figure*}
\centering
\includegraphics[width=12cm]{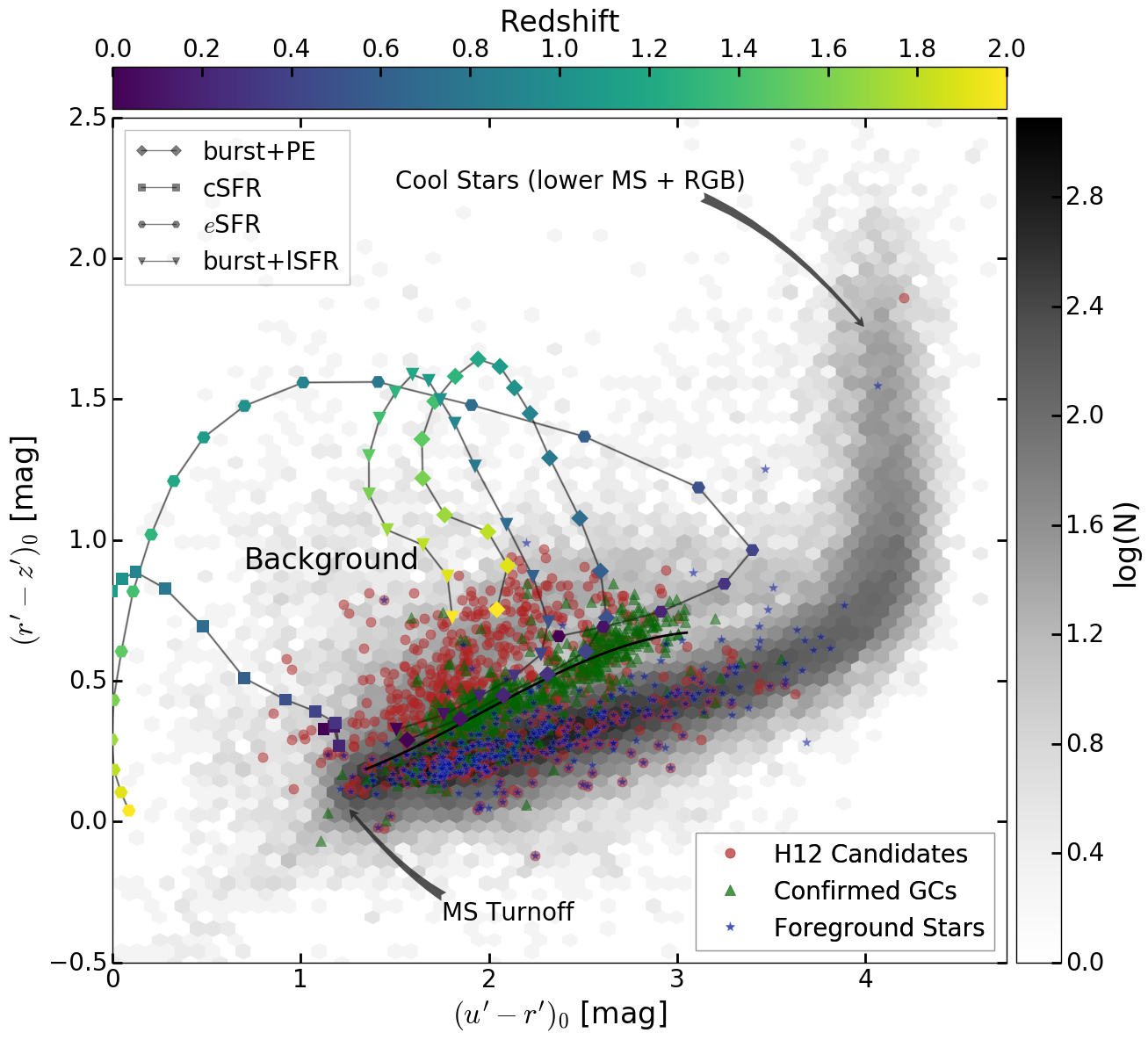}
\includegraphics[width=12cm]{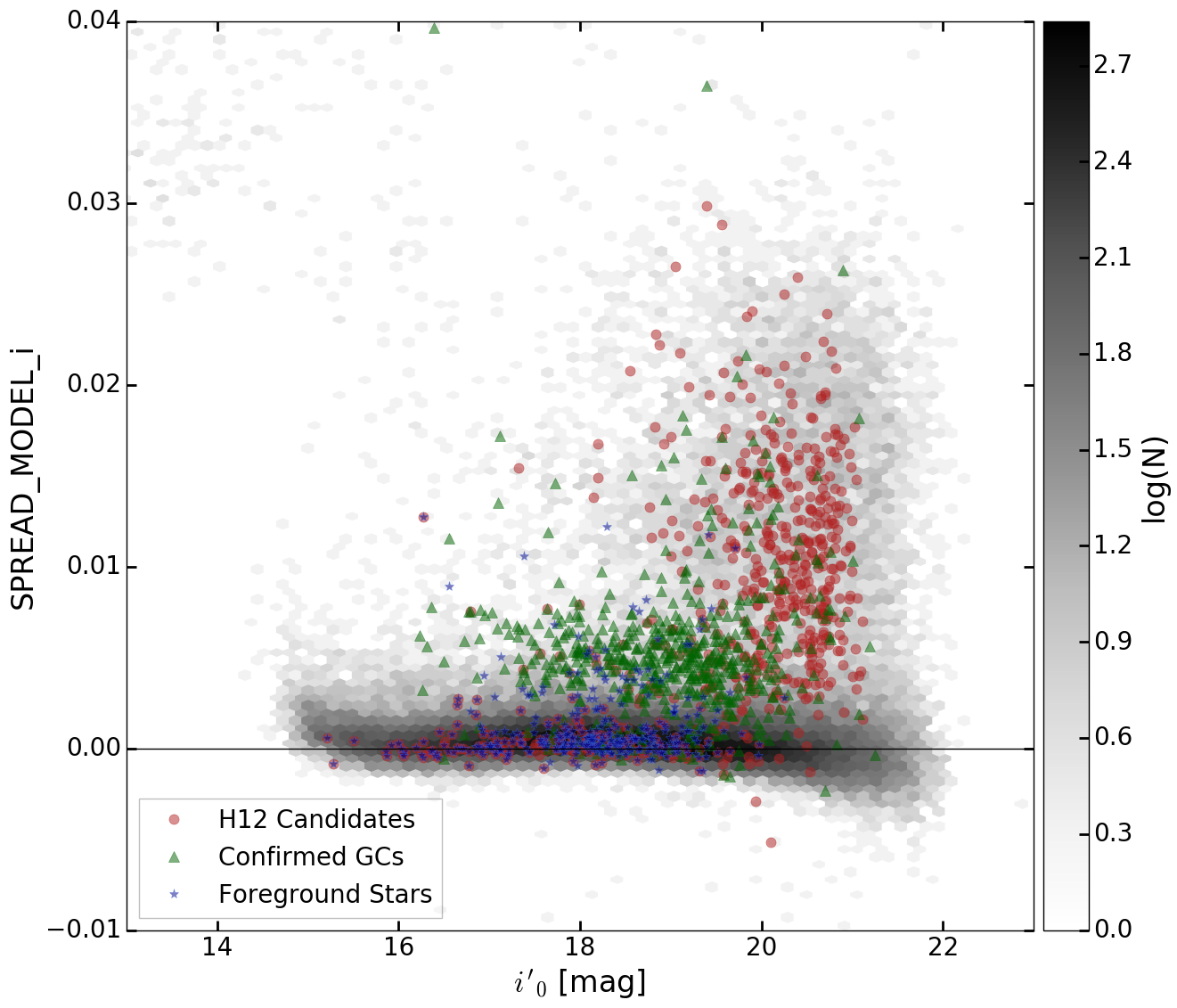}
\caption{The colour-colour and morphological GC selection diagrams. ({\it Top}): The $(r'-z')_0$ vs. $(u'-r')_0$ colours for all sources with complete $u'g'r'i'z'$ photometry in tile\,1 is shown (grey-scale number density map), with the H12 GC candidates (red dots), the confirmed population of GCs (green triangles), and known foreground stars (blue stars) over-plotted. The solid black curve represents a spline fit to the confirmed GCs and is meant to guide the eye. The large symbols mapped to the upper colour-bar show evolutionary tracks for the observed colours of a galaxy formed at $z\approx3$, assuming four different star formation histories (see \S\,\ref{sec:separation} for details on the SED modeling). ({\it Bottom}): The morphology-luminosity plane for the same populations (see \S\,\ref{sec:galcull}).}
\label{fig:urz}
\end{figure*} 

\subsection{Two-Colour GC Selection}
\label{sec:separation}
A common strategy for identifying GCs in galaxy systems outside of the Local Group is to exploit their relatively limited colour dispersions via colour-magnitude diagrams. For example, H12 used this technique to discover $\gtrsim800$ new GC candidates around \cena\ using $B$ and $R$ photometry. While this strategy is useful as a first guess, \cite{mun14} recently showed the dramatically increased effectiveness of identifying GCs in colour-colour diagnostic space covering the NUV to near-infrared (NIR) SED range. Presently, there is no matching wide-field NIR photometry available for \cena, so here we combine the $u'$, $r'$, and $z'$ filters from our survey which maximize the separation of GCs from background objects and, in particular, foreground stars.

The upper panel of Fig.\,\ref{fig:urz} illustrates the utility of leveraging the full optical SED for GC selection. The $(u'-r')_0$--$(r'-z')_0$ plane is shown for all 77\,336 sources detected in all five bands in tile\,1, with the logarithmic number density indicated by the grey-scaled hexagonal bins. We recover 691/833 GC candidates from the H12 catalogue, and over-plot them as red dots, while the 548/643 radial-velocity confirmed GCs \cite[][and references therein; Peng, E.~W., {\it private communication}]{woo10b} recovered in all five filters are plotted as green triangles. Finally, the blue stars in Fig.\,\ref{fig:urz} show the 369/373 confirmed foreground stars that were recovered \citep[][Peng, E.~W., {\it private communication}]{pen04a}.

The key to this selection technique lies in the inclusion of the NUV $u'$-band fluxes. While the effective temperatures of cool stars and red giant branch (RGB) populations are probed by the $(r'-z')_0$ colour, the $(u'-r')_0$ index straddles the $4000$\,\AA\ break and is thus sensitive to hot stellar populations in star-forming galaxies and/or hot horizontal branch (HHB) stars in GCs. The $u'$-band flux contributed by HHB stars in metal-poor GCs thus makes them appear bluer in $(u'-r')_0$ than foreground stars at a given $(r'-z')_0$, while GCs lacking an extended horizontal branch will not exhibit bluer colours and fall along the redder areas of the GC sequence that is clearly separate from the stellar locus.

The larger coloured symbols in the upper panel of Fig.\,\ref{fig:urz} show the redshift evolutionary tracks of the SED of a galaxy formed at $z\approx3$, based on the {\sc P\'egase} population synthesis models \citep{fio97}. Four star formation histories are assumed for the simulated galaxy, including an initial starburst followed by passive evolution (burst+PE), a constant star formation rate (cSFR), an exponentially declining SFR ($e$SFR), and a galaxy formed via an initial starburst followed by constant, low-level star formation (burst+lSFR). The pathways followed by the galaxies through the $u'r'z'$-plane differ significantly among themselves, and give rise to the diffuse background population towards redder $(r'-z')_0$ colours at a given $(u'-r')_0$. While the burst+PE model seen at low redshift closely follows the GCs, galaxies that experience virtually any residual star formation will remain satisfactorily distinct from GCs at all $z$.

Since a canonical stellar initial mass function \citep[e.g.\ Salpeter IMF or Chabrier IMF;][]{sal55,cha03} dictates that the stellar mass of a GC is dominated by cool, low-mass stars, their contribution to the overall light is weak compared to those near the main sequence turnoff (MSTO) and the RGB, leading to the clear separation of GCs from the plume of stars seen at redder $(r'-z')_0$, while being more similar at the MSTO. Altogether, the $(u'-r')_0$--$(r'-z')_0$ is the most effective optical two-colour diagnostic plane for GC selection. In practice, the upper panel of Fig.\,\ref{fig:urz} shows that the confirmed GCs populate a well-defined sequence that falls redward of the $(r'-z')_0$ foreground stellar locus, and are distinct from the MSTO of Galactic stars at redder colours. The reported H12 candidate sample is contaminated by a large number of likely false-positives, as many of them fall directly on the stellar locus (i.e.\ blueward of the GC sequence), or towards redder $(r'-z')_0$ colours where the population of confirmed GCs is lower, and the likelihood of background galaxy contamination increases.

We augment our photometric colour selection technique with morphological information as shown in the bottom panel of Fig.\,\ref{fig:urz}, and discussed in more detail in \S\,\ref{sec:galcull} (see also Fig.\,\ref{fig:galpoint}). This panel shows the {\sc Source Extractor} (SE) parameter {\sc spread\_model} against the apparent $i'$-band magnitude for the same sources and symbol definitions as above. A relatively clean separation can be seen between foreground stellar sources that show a median {\sc spread\_model}\ $=1.9\pm4.5\times10^{-4}$ and the confirmed GCs which show modestly larger median {\sc spread\_model}\ $=5.0\pm5.3\times10^{-3}$. Interestingly, the H12 sample shows indications of heavy foreground star and background source contamination, as many fall directly on the stellar {\sc spread\_model}$\simeq$0.0 locus, or show $i'$--{\sc spread\_model} combinations inconsistent with the radial-velocity confirmed GC sample. The large number of potential H12 GC imposters may be due to the use of a single $(B\!-\!R)_0$ colour, and we defer a quantification of this sample to \S\,\ref{sec:gcselect}. In any case, Fig.\,\ref{fig:urz} shows the utility of leveraging the full optical SED, combined with morphological information, which we use in the following to construct a new catalogue of GC candidates around \cena.

%%%%%%%%%%%%%%%%%%%%%%%%%%%%%%%
%%%%%%%%%%%%%%%%%%%%%%%%%%%%%%%
%%%%%%%%%%%%%%%%%%%%%%%%%%%%%%%

\subsection{Galaxy Culling}
\label{sec:galcull}
\begin{table*}
	\centering
	\caption{Source catalogue summary. Columns list tiles\,1--7 with the final column showing the totals of each type of source indicated by the rows, with all sources shown in the first row, point and GC-like sources in the second, and GC candidates in the bottom row.}
	\label{tbl:sources}
	\begin{tabular}{lrrrrrrrr}
		\hline\hline
		 			& 	tile\,1	&	tile\,2	&	tile\,3	&	tile\,4	&	tile\,5	&	tile\,6	&	tile\,7	&	Total	\\
		\hline
		$N_{\rm src}$	&	77\,336	&	67\,349	&	60\,627	&	64\,316	&	77\,528	&	87\,345	&	82\,348	&	516\,849	\\
		$N_{\rm pt}$	&	68\,434	&	58\,167	&	51\,336	&	51\,876	&	66\,780	&	74\,861	&	67\,495	&	438\,949	\\
		$N_{\rm GC}$	&	761		&	362		&	300		&	256		&	344		&	327		&	326		&	2\,676	\\
		\hline
	\end{tabular}\\
\end{table*}

The first step in cleaning the source catalogues of non GC-like objects is the removal of extended background sources. To this effect, we use the {\sc SE}/{\sc PSFEx} parameters {\sc spread\_model} and {\sc spreaderr\_model} following a procedure similar to previous works \citep[e.g.][]{des12,ann13,kop15}. {\sc spread\_model} is an improvement on the older {\sc class\_star} star--galaxy separation parameter. It uses the local PSF model, $\Phi$, to discriminate between star-like and more extended sources by convolving it with a circular exponential disk based on the PSF model FWHM and calculating,
\begin{equation}
\label{eq:spread}
\textsc{spread\_model}=\frac{\Phi^T{\bf x}}{\Phi^T\Phi}-\frac{{\bf G}^T{\bf x}}{{\bf G}^T\Phi}
\end{equation}
where ${\bf G}$ is the more extended model and ${\bf x}$ is the image vector centred on the source. In this way, {\sc spread\_model} forms an effective discriminant between the modelled PSF and a more extended source at bright magnitudes, with blending occurring at fainter magnitudes that is quantified by the related {\sc spreaderr\_model} parameter.

\begin{figure}
\centering
\includegraphics[width=\columnwidth]{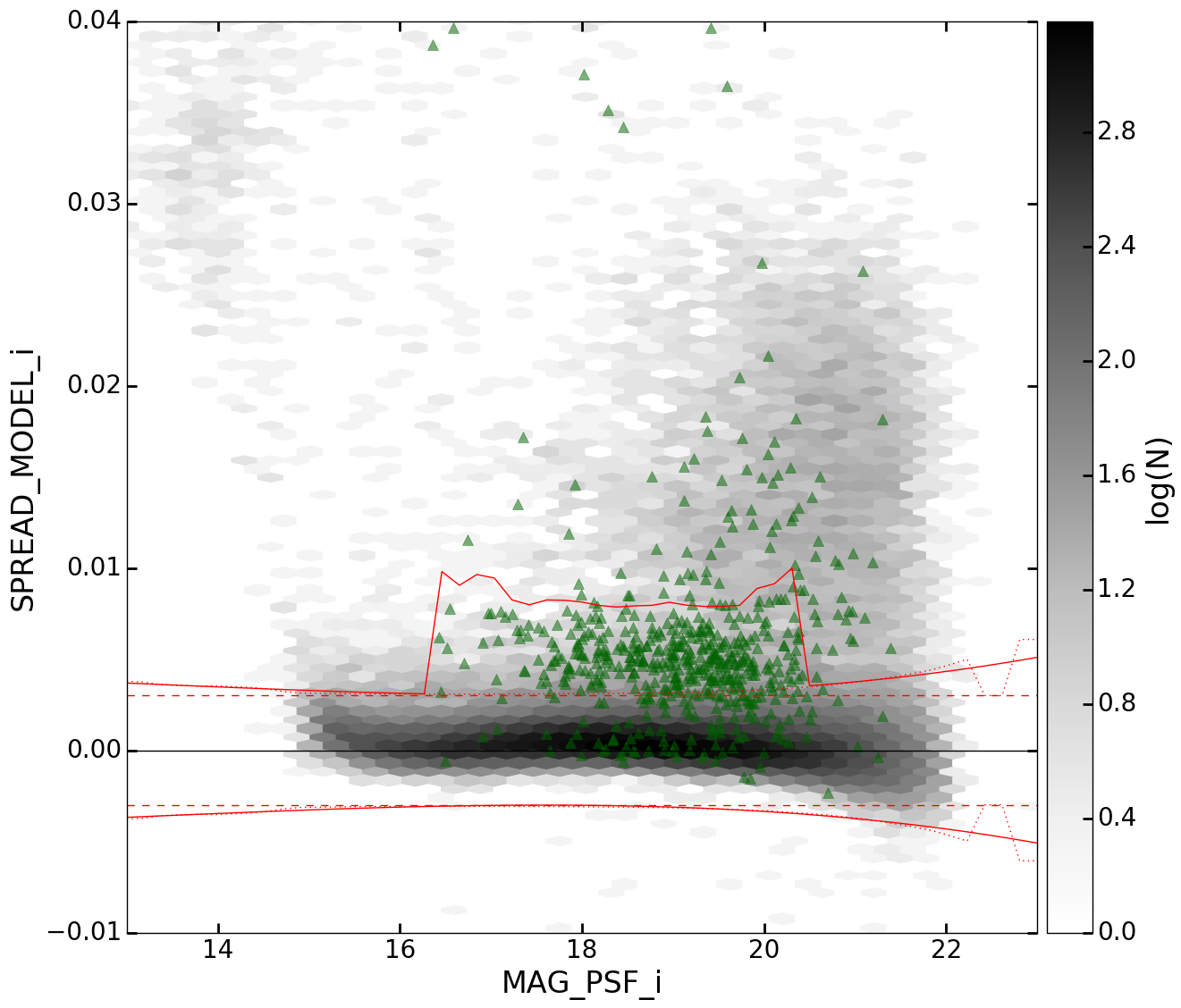}
\caption{The galaxy--point source classification diagram. As in the lower panel of Fig.\,\ref{fig:urz}, the logarithmic number density of all sources detected in tile\,1 is indicated by the grey-scaled hexagonal binning, and green triangles denote the confirmed GCs. Note that many of the brighter GCs are marginally resolved at the distance of \cena, indicated by the slight departures from {\sc spread\_model}$\simeq$0. The red curves indicate the fine-tuning criteria of separating point and GC-like sources from the extended background sources (see text).}
\label{fig:galpoint}
\end{figure}

Fig.\,\ref{fig:galpoint} shows this technique in practice, which shows the same data as the lower panel of Fig.\,\ref{fig:urz}, but without known stars or H12 candidates. Red curves show extra fine-tuning to account for the marginally resolved nature of GCs at the distance of \cena. True point sources are expected to show {\sc spread\_model}\ $=0.0\pm(0.003+${\sc spreaderr\_model}$)$ \citep{des12}, and we combine these criteria with an extra selection strategy to create our point and GC-like source catalogue. We note here that the term ``GC-like'' is used to refer to all marginally resolved sources, which are further refined in \S\,\ref{sec:gcselect}. While the dashed red curve in Fig.\,\ref{fig:galpoint} indicates lines of {\sc spread\_model}\ $=0.0\pm0.003$, we instead use the non-continuous, piecewise function,
\[ 
\mu\left(m_{i'}\right)= 
\begin{cases} 
	 0.003+\epsilon(m_{i'})				&	m_{i'}\leq 16.0 		\\
	 0.003+\epsilon(m_{i'})+\delta(m_{i'})	        &	16.0 < m_{i'} < 20.4	\\
	 0.003+\epsilon(m_{i'})				&	m_{i'}\geq 20.4 		\\
   \end{cases}
\]

Here, $m_{i'}$ is the apparent $i'$-band PSF magnitude, $\epsilon(m_{i'})$ is the running mean {\sc spreaderr\_model} along bins in $m_{i'}$, and $\delta(m_{i'})$ is the running median of {\sc spread\_model} for the known GCs within the same bins. We calculate $\delta(m_{i'})$ for $16.0\!<m_{i'}\!<20.4\,{\rm mag}$ to capture the bulk of the confirmed GC population, while avoiding saturated stars at brighter magnitudes, and minimizing contamination of extended background sources at faint magnitudes. To calculate $\delta(m_{i'})$, we first $2.3\sigma$-clip the binned GC data, and find that restricting the range to a relatively conservative $m_{i'}\!<\!20.4$\,mag severely limits the amount of potential background contamination at fainter magnitudes. The dashed red lines on Fig.\,\ref{fig:galpoint} show the $\pm0.003$ bounds on {\sc spread\_model}, and the dotted lines show the {\sc spread\_model}$\pm(0.003+\epsilon(m_{i'}))$ relation, to which we fit a third degree polynomial\footnote{{\sc Python} package: {\sc NumPy/poly1d}} and add $\delta(m_{i'})$ for {\sc spread\_model}$>0.0$ to obtain the solid red relations shown.

Using these criteria, we select all sources falling within the solid red curves in Fig.\,\ref{fig:galpoint}, successfully capturing 520/569 ($\sim90$ percent) of confirmed GCs, and repeat the process for tiles\,2--7 (see Fig.~\ref{fig:dither}) to construct our final point and GC-like source catalogues, including marginally resolved sources that fall within magnitudes typical of GCs. Using a list of 157 confirmed background galaxies in tile\,1 compiled using the {\it Nasa Extragalactic Database} (NED\footnote{\url{http://ned.ipac.caltech.edu}}), we find that this procedure successfully culls 140 objects ($\sim90\%$), with the remaining 17 having {\sc spread\_model} consistent with being unresolved. In general, we find that $\sim\!80\!-\!85\%$ of the sources in the SCABS imaging are point or GC-like, and we list their photometric measurements in Table\,\ref{tbl:point_cat} with their associated statistical and systematic error estimates listed in Table\,\ref{tbl:point_errs}. 

\begin{table*}
	\centering
	\caption{Catalogue of point and GC-like sources. Cols.\ (1) and (2) list on-sky coordinates, followed by their $u'$, $g'$, $r'$, $i'$, and $z'$ PSF magnitudes in cols.\ (3)--(7). Listed magnitudes are not corrected for foreground reddening. This table is available in its entirety in machine-readable form.}
	\label{tbl:point_cat}
	\begin{tabular}{cccccccccccc}
		\hline\hline
		$\alpha(J2000)$ 	& 	$\delta(J2000)$	&	$u'$		&	$g'$		&	$r'$		&	$i'$		&	$z'$		\\
		($hh:mm:ss$)		&	($^{\circ}$:$'$:$''$)	&	(mag)	&	(mag)	&	(mag)	&	(mag)	&	(mag)	\\
		\hline
		13:10:38.71	&	$-$42:17:52.21	&	19.82	&	18.40	&	17.89	&	17.89	&	17.91	\\
		13:10:38.73	&	$-$42:29:04.31	&	22.79	&	20.79	&	20.04	&	19.87	&	19.84	\\
		13:10:38.75	&	$-$42:22:30.00	&	19.76	&	17.44	&	16.58	&	16.55	&	16.50	\\
		13:10:38.91	&	$-$42:15:17.65	&	21.93	&	19.45	&	18.48	&	18.27	&	18.18	\\
		13:10:38.96	&	$-$42:21:27.92	&	17.93	&	16.57	&	15.92	&	15.39	&	15.42	\\
		13:10:39.08	&	$-$42:18:09.06	&	22.21	&	19.86	&	18.99	&	18.82	&	18.72	\\
		13:10:39.20	&	$-$42:14:16.45	&	21.52	&	19.97	&	19.42	&	19.32	&	19.40	\\
		13:10:39.33	&	$-$42:14:38.75	&	18.63	&	17.01	&	16.55	&	16.50	&	16.56	\\
		13:10:39.40	&	$-$42:14:34.73	&	20.12	&	18.99	&	18.55	&	18.45	&	18.47	\\
		13:10:39.48	&	$-$42:24:09.20	&	22.97	&	20.03	&	18.86	&	18.56	&	18.42	\\
		\hline\hline
	\end{tabular}
\end{table*}

\begin{table*}
	\centering
	\caption{Catalogue of point and GC-like source statistical and systematic photometric error budgets derived and described in detail in Paper\,I. The first two columns show coordinates corresponding to the source list in Table\,\ref{tbl:point_cat}, followed by sets of two columns that list their statistical and systematic error estimates for each filter. This table is available in its entirety in machine-readable form.}
	\label{tbl:point_errs}
	\begin{tabular}{cccccccccccc}
		\hline
		\hline
		$\alpha(J2000)$ 	& 	$\delta(J2000)$	&	$\delta u'_{\rm stat}$	&	$\delta u'_{\rm sys}$	&	$\delta g'_{\rm stat}$	&	$\delta g'_{\rm sys}$	&	$\delta r'_{\rm stat}$	&	$\delta r'_{\rm sys}$	&	$\delta i'_{\rm stat}$	&	$\delta i'_{\rm sys}$	&	$\delta z'_{\rm stat}$	&	$\delta z'_{\rm sys}$	\\
		($hh:mm:ss$)		&	($^{\circ}$:$'$:$''$)	&	(mag)			&	(mag)			&	(mag)			&	(mag)			&	(mag)				&	(mag)			&	(mag)			&	(mag)			&	(mag)			&	(mag)			\\
		\hline
		13:10:38.71	&	$-$42:17:52.21	&	0.01	&	0.08	&	0.01	&	0.03	&	0.01	&	0.08	&	0.01	&	0.02	&	0.01	&	0.02	\\
		13:10:38.73	&	$-$2:29:04.31	&	0.10	&	0.16	&	0.04	&	0.08	&	0.03	&	0.09	&	0.02	&	0.06	&	0.02	&	0.06	\\
		13:10:38.75	&	$-$42:22:30.00	&	0.01	&	0.08	&	0.01	&	0.03	&	0.01	&	0.08	&	0.01	&	0.01	&	0.01	&	0.01	\\
		13:10:38.91	&	$-$42:15:17.65	&	0.05	&	0.10	&	0.01	&	0.04	&	0.01	&	0.08	&	0.01	&	0.02	&	0.01	&	0.02	\\
		13:10:38.96	&	$-$42:21:27.92	&	0.01	&	0.08	&	0.01	&	0.03	&	0.01	&	0.08	&	0.01	&	0.01	&	0.01	&	0.01	\\
		13:10:39.08	&	$-$42:18:09.06	&	0.06	&	0.12	&	0.02	&	0.05	&	0.01	&	0.08	&	0.02	&	0.03	&	0.01	&	0.03	\\
		13:10:39.20	&	$-$42:14:16.45	&	0.03	&	0.09	&	0.02	&	0.05	&	0.02	&	0.09	&	0.02	&	0.04	&	0.01	&	0.05	\\
		13:10:39.33	&	$-$42:14:38.75	&	0.01	&	0.08	&	0.01	&	0.03	&	0.01	&	0.08	&	0.01	&	0.01	&	0.01	&	0.01	\\
		13:10:39.40	&	$-$42:14:34.73	&	0.01	&	0.08	&	0.01	&	0.03	&	0.01	&	0.08	&	0.01	&	0.02	&	0.01	&	0.03	\\
		13:10:39.48	&	$-$42:24:09.20	&	0.12	&	0.16	&	0.02	&	0.05	&	0.01	&	0.08	&	0.01	&	0.02	&	0.01	&	0.03	\\
		\hline
		\hline
	\end{tabular}
\end{table*}

%%%%%%%%%%%%%%%%%%%%%%%%%%%%%%%
%%%%%%%%%%%%%%%%%%%%%%%%%%%%%%%
%%%%%%%%%%%%%%%%%%%%%%%%%%%%%%%

\subsection{Star--Globular Cluster Classification} 
\label{sec:gcselect}
Having effectively culled extended background sources from the catalogues, the next step is to differentiate between star- and GC-like objects. Similar to previous work \citep[e.g.][]{pen04a}, we use the full optical SED coverage in colour-colour space to accomplish this task. While this colour-space is exploited to construct our final list of GC candidates, there is still some ambiguity between the stellar and GC loci toward bluer colours. The following describes a novel probabilistic approach to separate GCs from stars including efforts to quantify the level of confidence for each source we have in our final GC candidate catalogue.

\begin{figure}
\centering
\includegraphics[width=\linewidth]{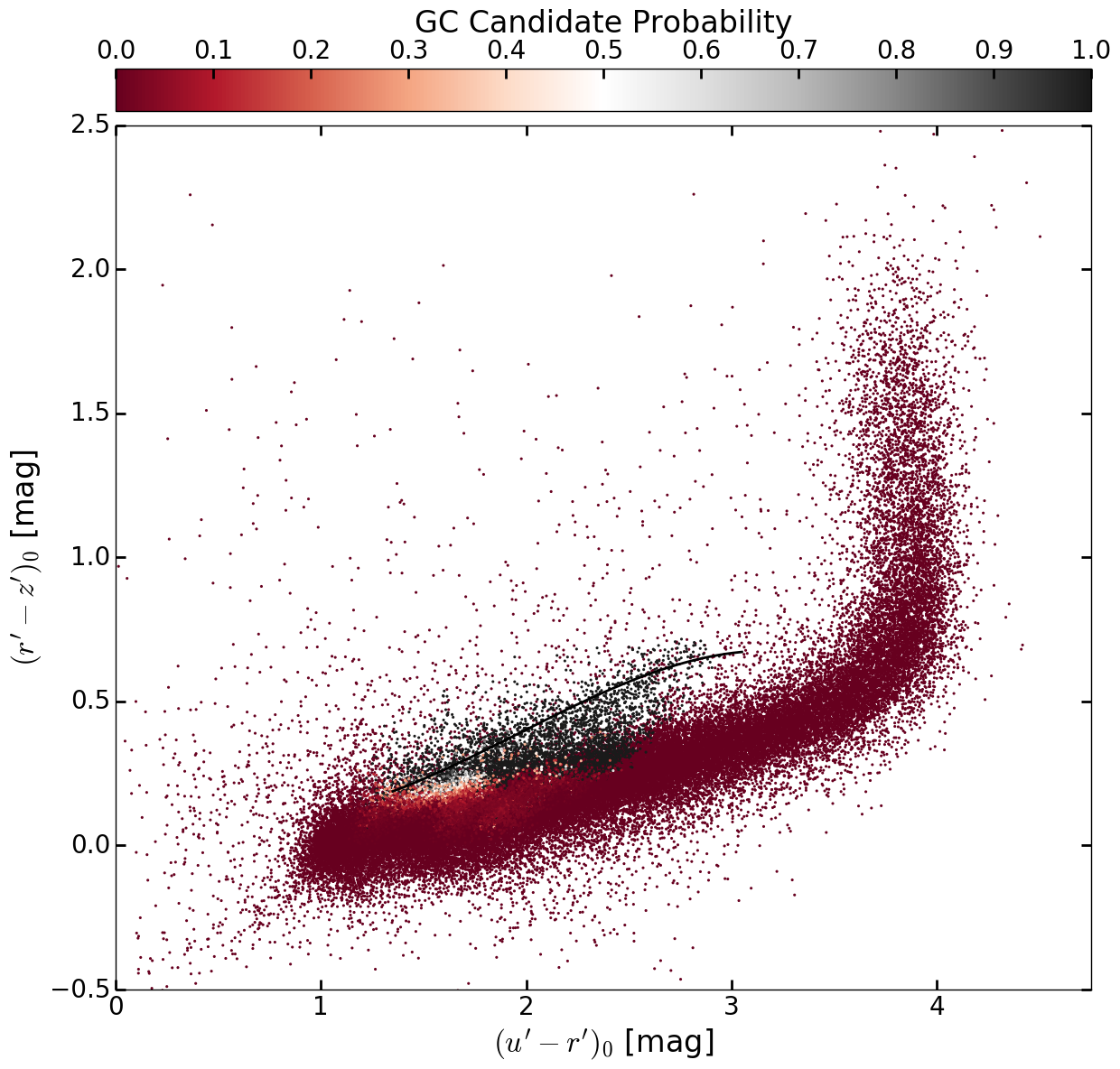}
\includegraphics[width=\linewidth]{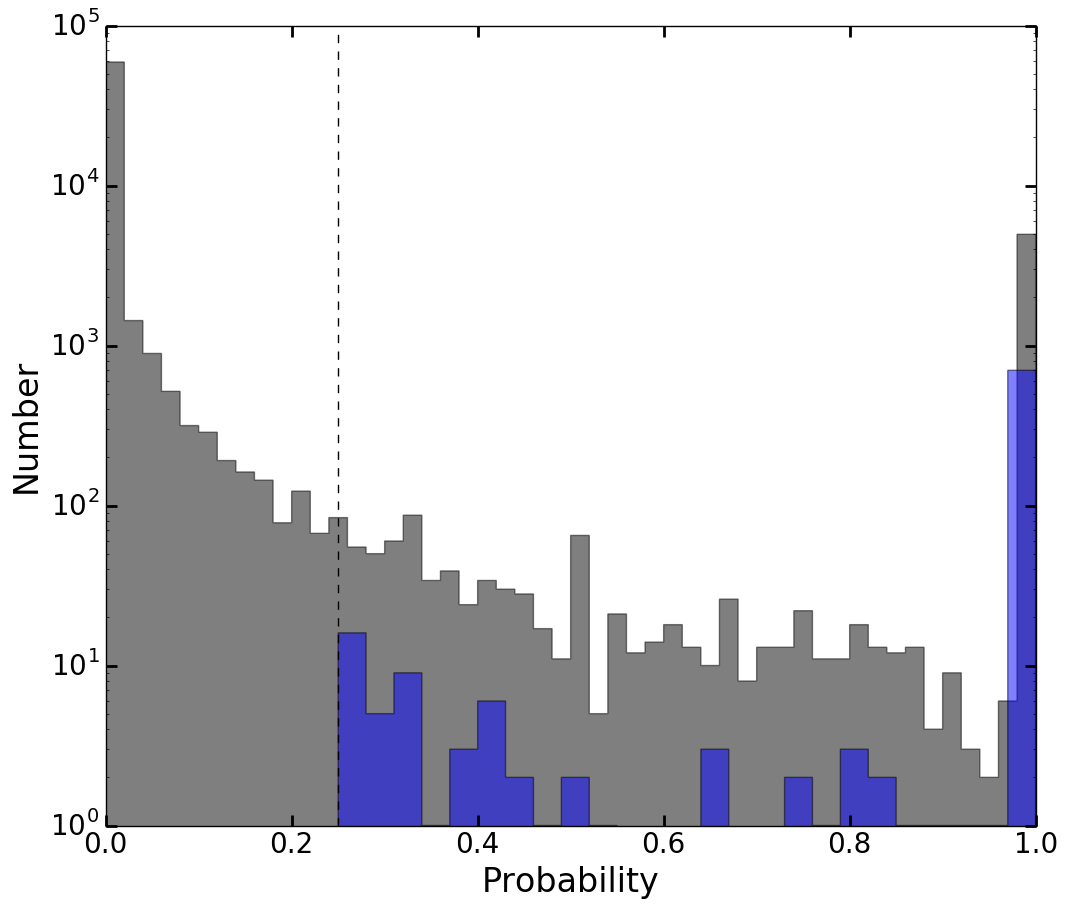}
\caption{GC candidate probabilities for all tile\,1 point- and GC-like sources. ({\it Top}): The $u'r'z'$-diagram with symbol colour parameterizing $P({\rm GC})$. $P({\rm GC})\!\approx\!1.0$ objects cluster around the GC locus (black curve) with a sharp gradient at bluer colours where the stellar sequence begins to dominate. ({\it Bottom}): The distribution of $P({\rm GC})$ for the whole sample (grey) and final GC candidates (blue). The black dashed line denotes the initial cut on $P({\rm GC})$ before further refinement is performed. The final list is dominated by high $P({\rm GC})$ candidates, with $\sim6$ percent of candidates showing $P({\rm GC})\leq0.5$.}
\label{fig:urzprobs}
\end{figure} 

\subsubsection{Modeling the Stellar and GC Colour-Colour Space}
We first generate a catalogue of 369 spectroscopically confirmed foreground stars from \cite{pen04a}, supplemented by newly confirmed foreground stars (Peng et al., {\it private communication}). From this catalogue, we model a population of foreground stars using the Besan\c{c}on model of stellar population synthesis of the Galaxy\footnote{\url{http://model.obs-besancon.fr}} \citep{rob03}. For each SCABS tile, we query the Besan\c{c}on model for differential counts of foreground stars in the range $15.0\leq m_{u'} \leq25.0$\,mag, considering only $(u'-g')_0$, $(g'-r')_0$, $(r'-i')_0$, and $(i'-z')_0$ colours in the range of the known stellar and GC populations. The queries are made over areas equivalent to the DECam footprint, centred on the coordinates of tiles\,1--7. The catalogue of known stars is then randomly drawn from until the expected population is reached, and assigned their corresponding $(u'-r')_0\pm\delta_{u'r'}$ and $(r'-z')_0\pm\delta_{r'z'}$ colours where $\delta$ is drawn from a uniform $U(0,\Delta)$ distribution and $\Delta$ is the {\sc SE} photometric error in a given band. In this way, for tiles\,1--7, the colour--colour space is populated by an expected population of foreground stars, based on photometry of known stellar sources.

Similarly, a model of GCs is constructed based on the 548 confirmed GCs with complete $u'g'r'i'z'$ photometry. As with the modelled foreground stars, we randomly draw from the sample of known GCs until the modelled population reaches an assumed total population of GCs. With no prior knowledge of the true population, we assume a maximum expected number of $N_{\rm pred}=2\,000$ based on previous population estimates for tile\,1 \citep{har84a,har02b,har06,har10b}, where the surface number density of GCs is presumed to be highest, and a correspondingly lower number for tiles\,2--7. For the latter regions, we assume $N_{\rm pred}=200$ to uniformly sample the $(u'-r')_0$--$(r'-z')_0$ parameter space and minimize stochastic under-sampling of the $u'r'z'$ diagram. 

\subsubsection{Deriving GC-Candidate Likelihoods}
Having populated the $u'r'z'$-diagram with modelled stars and GCs, the next step is to assign a probability for each object in the point source catalogue. For each source, a surrounding box is drawn based on its photometric errors. The number of modelled GCs within the selection box, $n_{\rm GC}$, as well as the number of modelled stars, $n_{*}$, are counted and used to calculate the probability of being a GC, $P({\rm GC})$ as,
\begin{equation}
\centering
P({\rm GC})=\frac{n_{\rm GC}}{n_{\rm GC}+n_{*}}
\end{equation}
and an initial rejection of any candidates with $P({\rm GC})<0.25$ is made. Fig.\,\ref{fig:urzprobs} illustrates the above procedure for all point-like sources in tile\,1. The top panel shows the $u'r'z'$-diagram for the sources with colour parameterizing $P({\rm GC})$. The same solid black relation as in Fig.\,\ref{fig:urz} follows the GC locus which is made visible by the cloud of $P({\rm GC})\simeq 1.0$ candidates with a sharp truncation in $P({\rm GC})$ toward bluer colours where the stellar sequence begins to dominate. The lower panel of Fig.\,\ref{fig:urzprobs} shows the distribution of $P({\rm GC})$ where the same transition from foreground stars to GC candidates is seen to flatten out for $P({\rm GC})\gtrsim0.25$, and sharply transitions to the $P({\rm GC})\simeq1.0$ population. Nonetheless, we note in the upper panel, the cloud of $P({\rm GC})\approx1.0$ points toward redder GC colours that stretches well below the locus of known GCs. For these objects, an extra morphological cut to filter out the most point-like ({\sc spread\_model}\,$\leq0.002$) sources removes virtually all of them, which suggests that they are likely to be primarily stars that are not well-sampled by our confirmed sample.

To summarize the selection state of affairs at this point in the procedure, we mention that, for tile\,1 where the surface number density of GCs is highest, the colour and $P({\rm GC})$ selection procedure serves to cull $\gtrsim90$ percent of point-like sources, with an additional $\sim80$ percent removed due to truly point source morphologies. For tiles\,2-7, where the surface number density of GCs is expected to be dramatically lower, we find a correspondingly, and consistently, higher colour/$P({\rm GC})$ cull rate of $\sim98$ percent, with an additional $\sim65-80$ percent removed on the second {\sc spread\_model} filtering.

\subsubsection{Understanding the Population Size Stochasticity}
We investigate the effects of changing the assumed total population, $N_{\rm pred}$, and show the results in Fig.\,\ref{fig:gc_poptest}. The GC selection procedure is run for the central tile (see Fig.~\ref{fig:dither}) and one of the outer tiles, but with the assumed true population of GCs, $N_{\rm pred}$, drawn from $U(10,4\,000)$ and $U(10,2\,000)$ distributions, respectively. The resulting $N_{\rm GC}$ for the individual iterations are shown by the small dots, which give rise to the binned averages (connected dots) braced by the $\pm1\sigma$ curves. We find that for tile\,1, $N_{\rm GC}$ climbs sharply for increasing $N_{\rm pred}\lesssim1\,500$, above which it remains relatively stable with a slowly increasing trend up to $N_{\rm pred}=4\,000$. For the Outer Ring, where the GC density is presumed to be dramatically lower, $N_{\rm GC}$ shows a relatively stable plateau for $200\lesssim N_{\rm pred}\lesssim800$, followed by a slowly increasing trend. In both cases, the positive $N_{\rm GC}$--$N_{\rm pred}$ correlation arises from the stochastic sampling of the GC locus. For example, if $N_{\rm pred}$ is assumed to be too small, then the modelled GC locus does not adequately sample the full $u'r'z'$ colour space, resulting in smaller $N_{\rm GC}$. On the other hand, particularly for tiles\,2-7, if one assumes too high $N_{\rm pred}$, then the modelled population risks oversampling regions that cut into the stellar locus, with $N_{\rm GC}$ larger than one might expect for the extreme halo of a gE. Given this result, we move forward with $N_{\rm pred}=2000$ and $200$, and account for potential stochastic sampling using an iterative strategy before building our final candidate catalogue as described below.

\subsubsection{Building the GC Candidate Catalog}
Star clusters of stellar masses, ${\cal M}_*\lesssim10^4\,M_\odot$ typically disperse on Gyr timescales \citep[see e.g.,][for a review]{por10}, whereas ${\cal M}_*\gtrsim2\times10^6\,M_\odot$ compact stellar systems enter the realm of ultra-compact dwarf galaxies (UCDs), where chemical and photometric properties begin to diverge from typical GCs \citep[e.g.][]{hac05,mie06,mie08a,tay10,tay15}. Given typical stellar $V$-band mass-to-light ratios for Local Group GCs of $\Upsilon_*^V\simeq2.0\,M_\odot L_\odot^{-1}$ \citep{mcl00,mcl08,str11}, this range in ${\cal M}_*$ translates to $V$-band luminosities of $0.5\times10^4\lesssim L_{V}/L_\odot \lesssim 10^6$. At a distance modulus of $(m-M)_0=27.88\,{\rm mag}$, these luminosities translate to apparent $V$-band magnitudes of $17.7\lesssim m_V/{\rm mag}\lesssim23.5$. With this in mind, we remove candidates outside of this range in $m_V$, assuming a transformation to $V$-band from $g'$ and $r'$ of \cite{jes05}:
\begin{equation}
\label{eq:grvconvert}
V=g'-0.58(g'-r')-0.01
\end{equation}

\begin{figure}
\centering
\includegraphics[width=\linewidth]{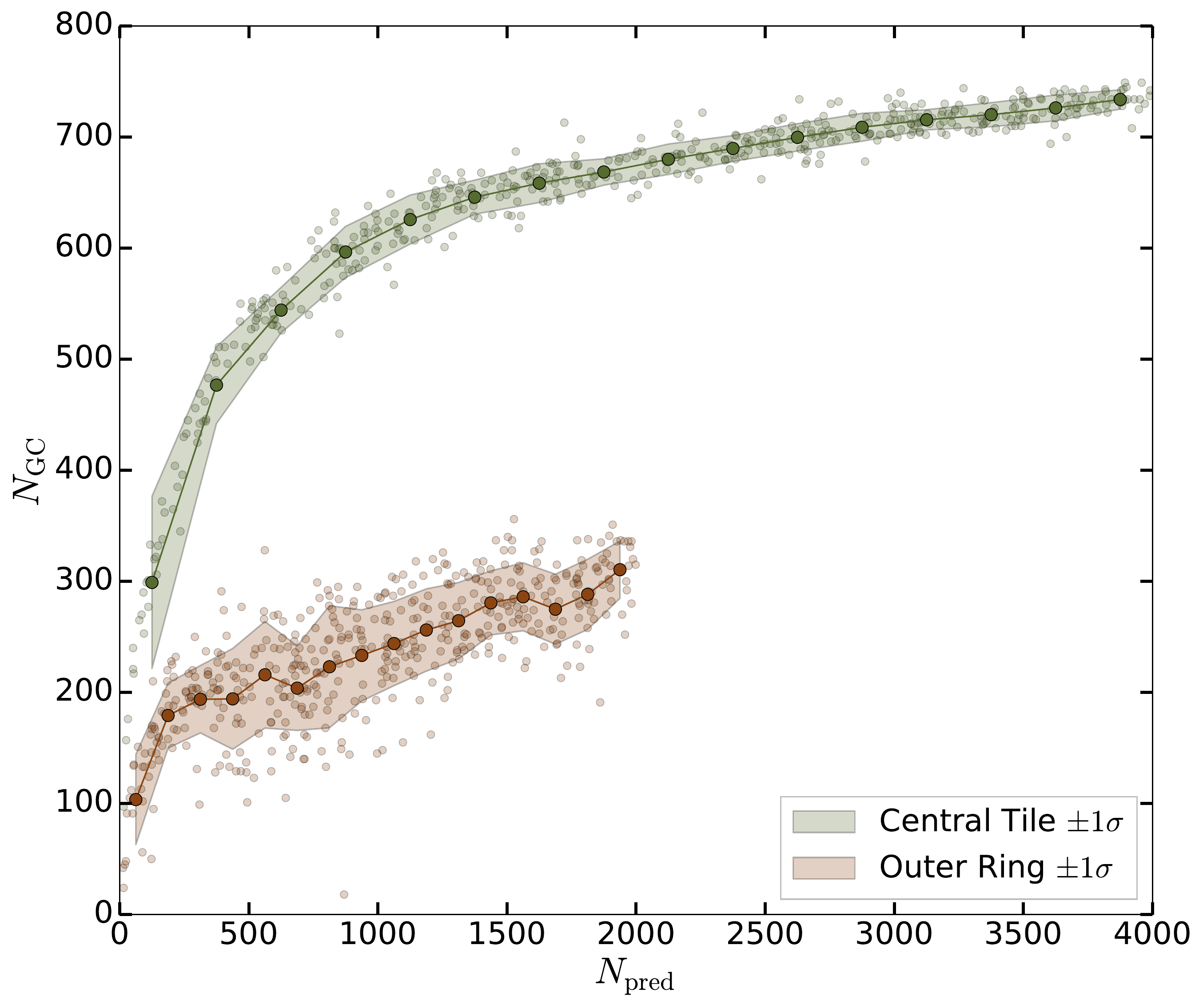}
\caption{Variation in predicted GC candidates with {\it a priori} assumed populations. The GC selection technique is applied to the data with several dozen realizations, each assuming a different true population of GCs. Assumed populations are drawn from $U(10,4000)$ for tile\,1 (green), reflecting the higher number density of GCs around \cena\ itself, and from $U(10,2000)$ for the Outer Ring. Smaller dots show the individual outcomes, while larger connected points and shading show the mean, binned $N_{\rm GC}\pm1\sigma$ results. $N_{\rm GC}$ is seen to generally stabilize for $N_{\rm pred}\gtrsim1\,500$ for the Central tile, while a temporary plateau appears for the Outer Ring at $200\lesssim N_{\rm pred}\lesssim800$.}
\label{fig:gc_poptest}
\end{figure}

\begin{figure}
\centering
\includegraphics[width=\linewidth]{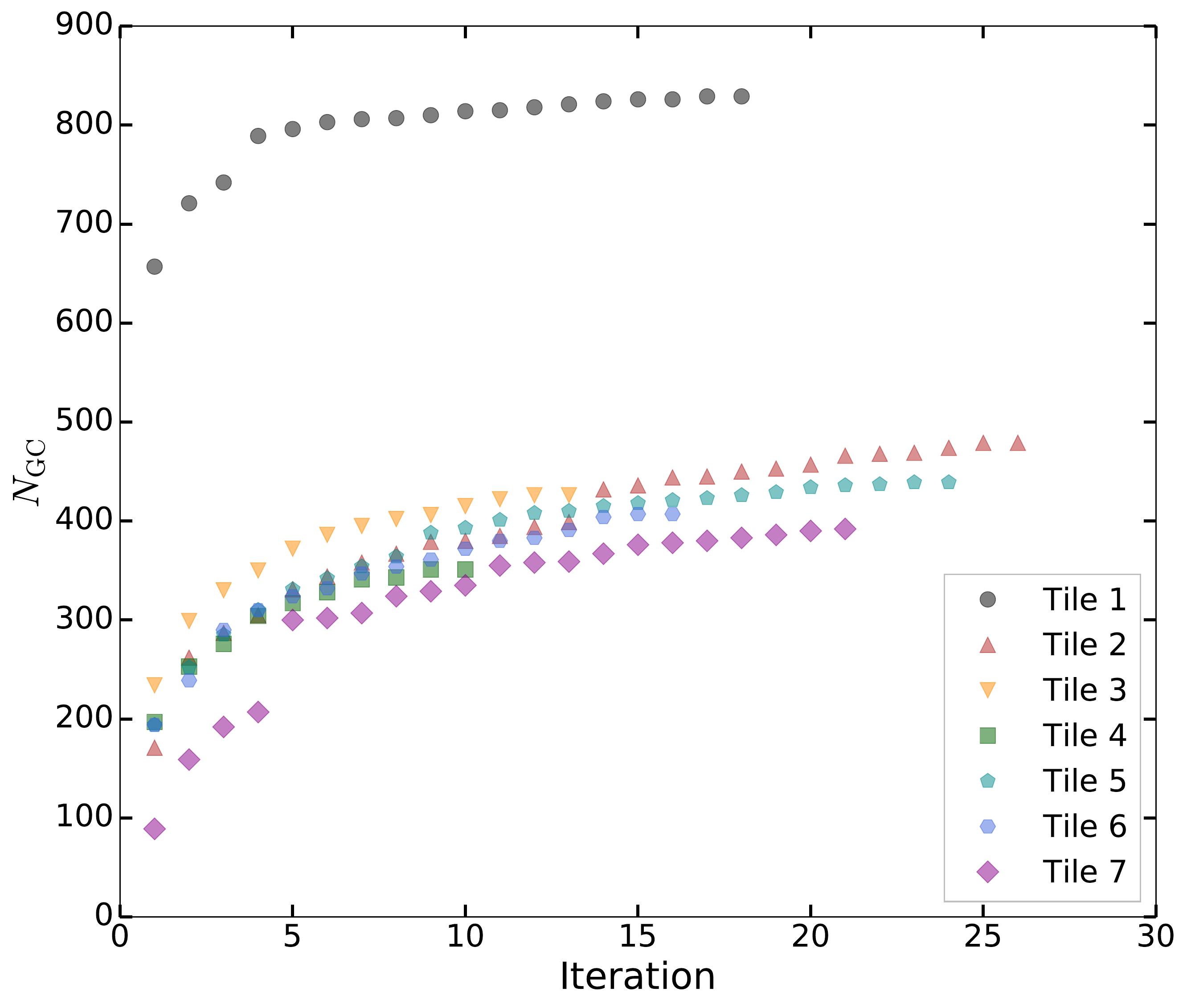}
\caption{An illustration of the iterative procedure undertaken to compile the final GC candidate catalogues. The ordinate shows the number of candidates identified in a given tile for each iteration indicated along the abscissa, with individual tiles shown by the different symbols/colours. Asymptotic behaviours are seen for all tiles, which all required several iterations before converging on their final values.}
\label{fig:iterate}
\end{figure}

Despite the robust GC selection procedure, there is some stochasticity remaining from the random nature of the stellar and GC modelling procedure. In other words, in a given pass through the algorithm, sub-regions in $u'r'z'$-space may be over- or under-sampled by the modelled GC and stellar populations, resulting in a handful of GC candidates that may be identified in one pass, but fail to be selected in another. To account for this effect, we iterate the algorithm over each tile, and keep unique candidates in each iteration until we reach a pass that fails to identify any new GCs. Fig.\,\ref{fig:iterate} illustrates this procedure for tiles\,1--7, with the growth of the final catalogues shown for each iteration. The results for each tile are indicated by the different symbols/colours, which rapidly flatten as the final $N_{\rm GC}$ are approached in typically $\lesssim20$ iterations. While a strict asymptotic limit is technically not reached, the results show that few new GC candidates would be identified with further iterations. We note that testing different assumed $N_{\rm pred}$ (Fig.\,\ref{fig:gc_poptest}) shows that the asymptotic behaviour persists regardless of assumed underlying populations, but with faster convergence accompanying larger assumed GC populations.

Finally, we find some sources with large photometric errors relative to their magnitudes, often spatially corresponding to the DECam chip-gaps, are found to contaminate the final GC candidate catalogues. These sources are $\kappa\sigma$-clipped out of the final samples, with $\kappa$ tuned for each tile until the scatter in magnitude error versus magnitude relations are minimized in all filters, resulting in the slightly smaller total candidate counts shown in Table\,\ref{tbl:sources} compared to those indicated on Fig.\,\ref{fig:iterate}. This last cut serves to typically reduce the final candidate lists by up to $25$ percent.

\begin{figure}
\centering
\includegraphics[width=\linewidth]{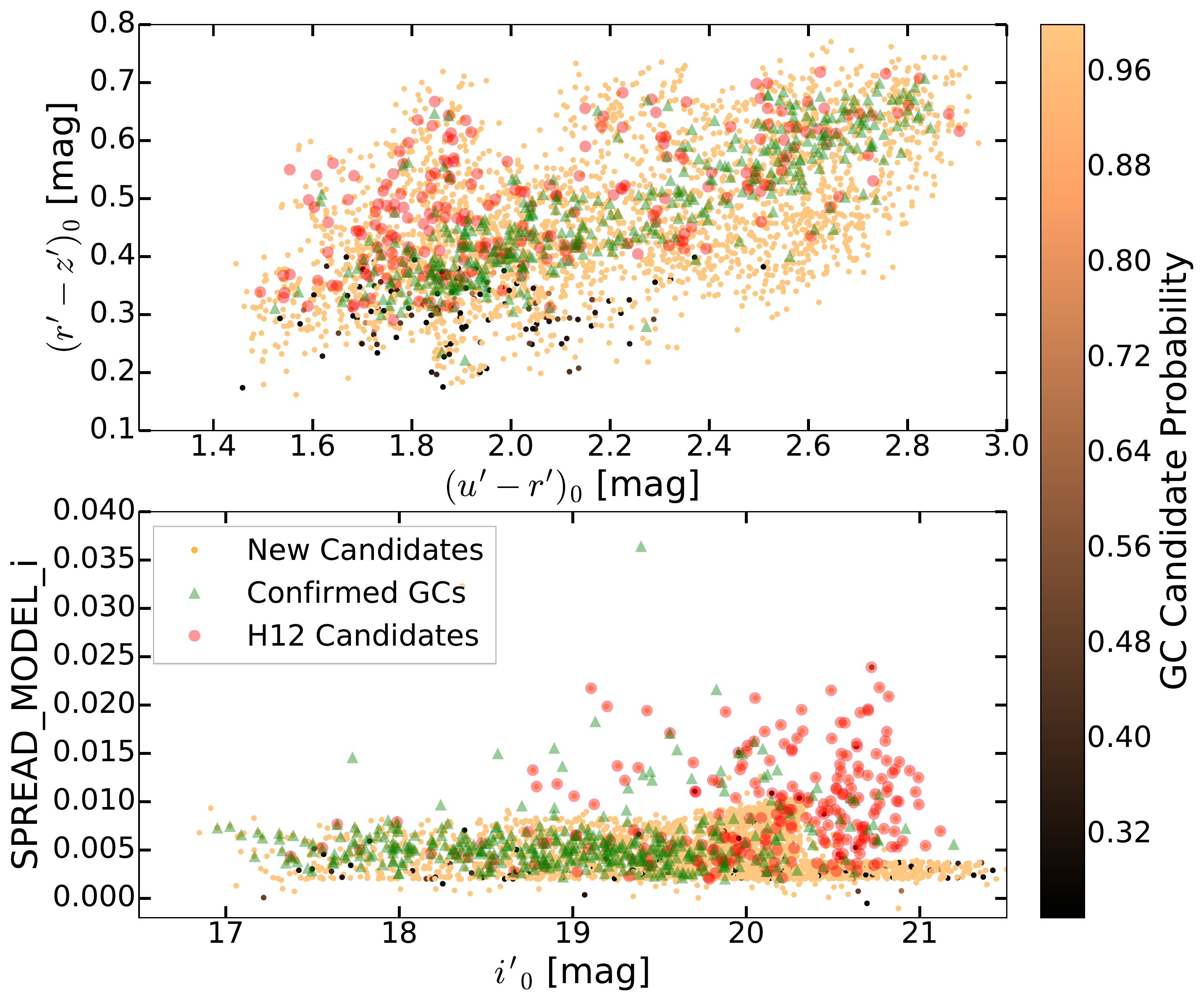}
\caption{The colour-colour and morphological diagrams for the new and previously reported GCs and candidates. As in Fig.\,\ref{fig:urz}, green triangles and red circles show confirmed GCs and H12 candidates, respectively, that survive our colour-selection criteria, while dots represent the new GC candidates, which are colour coded to indicate $P({\rm GC})$.}
\label{fig:urz_spread}
\end{figure} 

%%%%%%%%%%%%%%%%%%%%%%%%%%%%%%%
%%%%%%%%%%%%%%%%%%%%%%%%%%%%%%%
%%%%%%%%%%%%%%%%%%%%%%%%%%%%%%%
\begin{table*}
	\centering
	\caption{Catalogue of new GC candidates. Col.\ (1) lists new, homogenized identifications, while col.\ (2) lists former IDs and/or those reported in H12 (arbitrarily numbered). Cols.\ (3) and (4) list on-sky coordinates, with GC ``probabilities'', galactocentric radii, and azimuthal angles in cols.\ (5)--(7). Cols.\ (8)--(17) list $u'$, $g'$, $r'$, $i'$, and $z'$ PSF magnitudes without accounting for foreground extinction.}
	\label{tbl:gccand_cat}
	\begin{tabular}{cccccccccccc}
		\hline
		\hline
		ID	&	ID	&	$\alpha(J2000)$ 	& 	$\delta(J2000)$			&	$P_{\rm GC}$	&	$R_{\rm gc}$	&	$\Phi$		&	$u'$		&	$g'$	&	$r'$	&	$i'$	&	$z'$	\\	
		(T17)	&	(W10/H12	)	&	($hh:mm:ss$)		&	($^{\circ}$:$'$:$''$)	&				&	(arcmin)		&	($^\circ$ EoN)	&	(mag)	&	(mag)		&	(mag)	&	(mag)		&	(mag)	\\
		\hline
		T17-0001	&	...	&	13:10:45.60	&	$-$42:18:22.12	&	1.00	&	224.61	&	278.16	&	21.73	&	19.83	&	19.02	&	18.54	&	18.41	\\
		T17-0002	&	...	&	13:10:47.07	&	$-$42:24:37.20	&	1.00	&	223.15	&	276.98	&	23.25	&	21.82	&	20.93	&	20.35	&	20.20	\\
		T17-0003	&	...	&	13:10:57.17	&	$-$42:16:13.35	&	1.00	&	222.20	&	278.69	&	23.88	&	22.19	&	21.49	&	21.09	&	20.93	\\
		T17-0004	&	...	&	13:10:59.90	&	$-$42:28:18.49	&	1.00	&	219.40	&	276.37	&	22.99	&	21.64	&	20.96	&	20.49	&	20.39	\\
		T17-0005	&	...	&	13:11:00.02	&	$-$42:26:31.39	&	1.00	&	219.64	&	276.72	&	23.93	&	22.35	&	21.46	&	20.97	&	20.90	\\
		T17-0006	&	...	&	13:11:05.71	&	$-$42:16:11.52	&	1.00	&	220.12	&	278.78	&	22.92	&	21.45	&	20.84	&	20.43	&	20.20	\\
		T17-0007	&	...	&	13:11:17.95	&	$-$42:15:18.39	&	1.00	&	217.31	&	279.08	&	23.58	&	21.67	&	20.81	&	20.35	&	20.27	\\
		T17-0008	&	...	&	13:11:26.58	&	$-$42:20:48.01	&	1.00	&	214.10	&	278.07	&	21.29	&	19.68	&	19.06	&	18.66	&	18.62	\\
		T17-0009	&	...	&	13:11:26.86	&	$-$42:23:27.55	&	1.00	&	213.54	&	277.54	&	23.50	&	21.73	&	20.54	&	20.06	&	19.88	\\
		T17-0010	&	...	&	13:11:39.28	&	$-$42:36:19.89	&	1.00	&	208.56	&	275.04	&	23.19	&	22.04	&	21.44	&	20.86	&	20.96	\\
		\hline
		\hline
	\end{tabular}
\end{table*}

\begin{table*}
	\centering
	\caption{GC candidate photometric errors. The first three columns show IDs and coordinates corresponding to the source list in Table\,\ref{tbl:gccand_cat}, followed by sets of two columns that list the statistical and systematic error estimates for each filter, as described in \S\,\ref{sec:analysis}.}
	\label{tbl:gccand_err}
	\begin{tabular}{ccccccccccccc}
		\hline
		\hline
		ID	&	$\alpha(J2000)$ 	& 	$\delta(J2000)$		&	$\delta u'_{\rm stat}$	&	$\delta u'_{\rm sys}$	&	$\delta g'_{\rm stat}$	&	$\delta g'_{\rm sys}$	&	$\delta r'_{\rm stat}$	&	$\delta r'_{\rm sys}$	&	$\delta i'_{\rm stat}$	&	$\delta i'_{\rm sys}$	&	$\delta z'_{\rm stat}$		&	$\delta z'_{\rm sys}$	\\	
		(T17)	&	($hh:mm:ss$)		&	($^{\circ}$:$'$:$''$)	&	(mag)			&	(mag)			&	(mag)			&	(mag)			&	(mag)			&	(mag)			&	(mag)			&	(mag)			&	(mag)				&	(mag)			\\
		\hline
		T17-GC0001	&	13:10:45.60	&	$-$42:18:22.12	&	 0.01	&	 0.08	&	 0.01	&	 0.04	&	 0.01	&	 0.08	&	 0.01	&	 0.02	&	 0.00	&	 0.02	\\
		T17-GC0002	&	13:10:47.07	&	$-$42:24:37.20	&	 0.04	&	 0.13	&	 0.03	&	 0.11	&	 0.02	&	 0.10	&	 0.02	&	 0.06	&	 0.02	&	 0.07	\\
		T17-GC0003	&	13:10:57.17	&	$-$42:16:13.35	&	 0.08	&	 0.17	&	 0.05	&	 0.13	&	 0.04	&	 0.12	&	 0.03	&	 0.10	&	 0.04	&	 0.11	\\
		T17-GC0004	&	13:10:59.90	&	$-$42:28:18.49	&	 0.03	&	 0.11	&	 0.03	&	 0.09	&	 0.02	&	 0.10	&	 0.02	&	 0.06	&	 0.02	&	 0.08	\\
		T17-GC0005	&	13:11:00.02	&	$-$42:26:31.39	&	 0.08	&	 0.17	&	 0.06	&	 0.14	&	 0.04	&	 0.12	&	 0.03	&	 0.08	&	 0.04	&	 0.10	\\
		T17-GC0006	&	13:11:05.71	&	$-$42:16:11.52	&	 0.03	&	 0.11	&	 0.03	&	 0.08	&	 0.02	&	 0.10	&	 0.02	&	 0.06	&	 0.02	&	 0.07	\\
		T17-GC0007	&	13:11:17.95	&	$-$42:15:18.39	&	 0.06	&	 0.14	&	 0.03	&	 0.09	&	 0.02	&	 0.10	&	 0.02	&	 0.06	&	 0.02	&	 0.07	\\
		T17-GC0008	&	13:11:26.58	&	$-$42:20:48.01	&	 0.02	&	 0.08	&	 0.01	&	 0.03	&	 0.01	&	 0.08	&	 0.01	&	 0.02	&	 0.01	&	 0.03	\\
		T17-GC0009	&	13:11:26.86	&	$-$42:23:27.55	&	 0.05	&	 0.14	&	 0.03	&	 0.09	&	 0.02	&	 0.09	&	 0.01	&	 0.05	&	 0.01	&	 0.05	\\
		T17-GC0010	&	13:11:39.28	&	$-$42:36:19.89	&	 0.04	&	 0.12	&	 0.04	&	 0.13	&	 0.04	&	 0.12	&	 0.03	&	 0.08	&	 0.04	&	 0.11	\\
		\hline
		\hline
	\end{tabular}
\end{table*}

\subsection{Comparison with Previous Results}
\label{sec:compare}
The total GC candidate numbers ($N_{\rm GC}$) in Table\,\ref{tbl:sources} represent all survivors of the selection procedure described above, including those already known in the literature. Running the catalogue of \cite{woo10b} through our selection procedure results in 230/548 surviving GCs (a $\sim$60 percent culling fraction), while we can only say with confidence that 232/691 of the H12 candidates are truly GCs (a $\sim$67 percent cull). Cross matching the 761 GC candidates in tile\,1 with the 643 confirmed GCs \citep[][Peng et al., {\it private communication}]{woo10b} reveals 251 recovered clusters, with an additional 21 overlapping the surviving H12 candidate list. Subtracting these 272 candidates from our catalogue leaves a total of 2\,404 new GC candidates, a subsample of which is listed in Tables~\ref{tbl:gccand_cat} and \ref{tbl:gccand_err} along with spatial information and photometric measurements.

While it is beyond the scope of this work to determine the true nature of those H12 candidates that did not survive our selection procedure\tr{\footnote{This analyze the nature of these object will be part of subsequent spectroscopic studies.}}, it is worth mentioning two ``clumps'' of GC candidates reported by H12 between azimuthal angles $\Phi\approx200-220^\circ$, and at galactocentric radii $R_{\rm gc}=20$ and $40$\,kpc. Our final catalogue of likely GCs shows a slight enhancement of GC candidates corresponding to the ``clump'' nearest to \cena, but no compelling over-density is seen at $R_{\rm gc}\approx40$\,kpc. As noted by the authors, these ``clumps'' are likely to be background galaxy clusters that appear in their $B+R$ imaging as GCs, and further illustrates the importance of wide SED sampling when using colours to search for GCs. 

Cross-matching the 232 likely H12 GCs with the confirmed GCs leaves 199 unique candidates. Considering these, the 643 previously confirmed GCs, and our 2\,404 new candidates implies that a total approaching $N_{GC,t}\approx3\,200$ GCs may be present within $\sim140$\,kpc of \cena. This number is higher than the total population of 1000--2000 predicted in previous works \citep[e.g.][]{har84a,har04b,har06,har10b}, but we note that these works consider a markedly smaller spatial scale for \cena's halo than we examine in this work. In fact, if only the GCs and candidates within a projected $50\arcmin$ ($55\,{\rm kpc}$) of \cena\ are considered, outside of which there are indications of a transition to a different GC population (see \S\,\ref{sec:spatial_distros}), then $N_{GC,t}$ drops to $\sim1\,100$, in agreement with previous estimates. With that said, a more detailed discussion of the spatial and colour distributions is deferred to \S\ref{sec:spatial_distros}.

The new GC candidates are shown with previously confirmed or reported GCs in Fig.\,\ref{fig:urz_spread}. The top panel shows the $u'r'z'$ colour-colour diagram with colour illustrating $P({\rm GC})$ for the new GC candidates, while the confirmed GCs and surviving H12 candidates shown again as green triangles, and red dots. The bottom panel shows the same samples in the morphological classification diagram. The transition toward lower $P({\rm GC})$ can be seen toward bluer colours, and as constructed, the swarm of new GC candidates closely approximates the sequence of confirmed GCs. We note in the top panel that there are a handful of GCs that exhibit {\sc spread\_model} parameters $<0.002$. These objects correspond to confirmed GCs that did not survive the colour selection criteria, but were matched in the overall point-like source catalogues and are thus bona-fide GCs. The lack of large numbers is encouraging, as it suggests that we are not losing significant numbers from our final foreground star culling morphological cut.

%%%%%%%%%%%%%%%%%%%%%%%%%%%%%%%%%%%%%%%%%%%%%%%%%%
%%%%%%%%%%%%%%%%%%%%%%%%%%%%%%%%%%%%%%%%%%%%%%%%%%
%%%%%%%%%%%%%%%%%%%%%%%%%%%%%%%%%%%%%%%%%%%%%%%%%%
\begin{figure*}
\centering
\includegraphics[width=\linewidth]{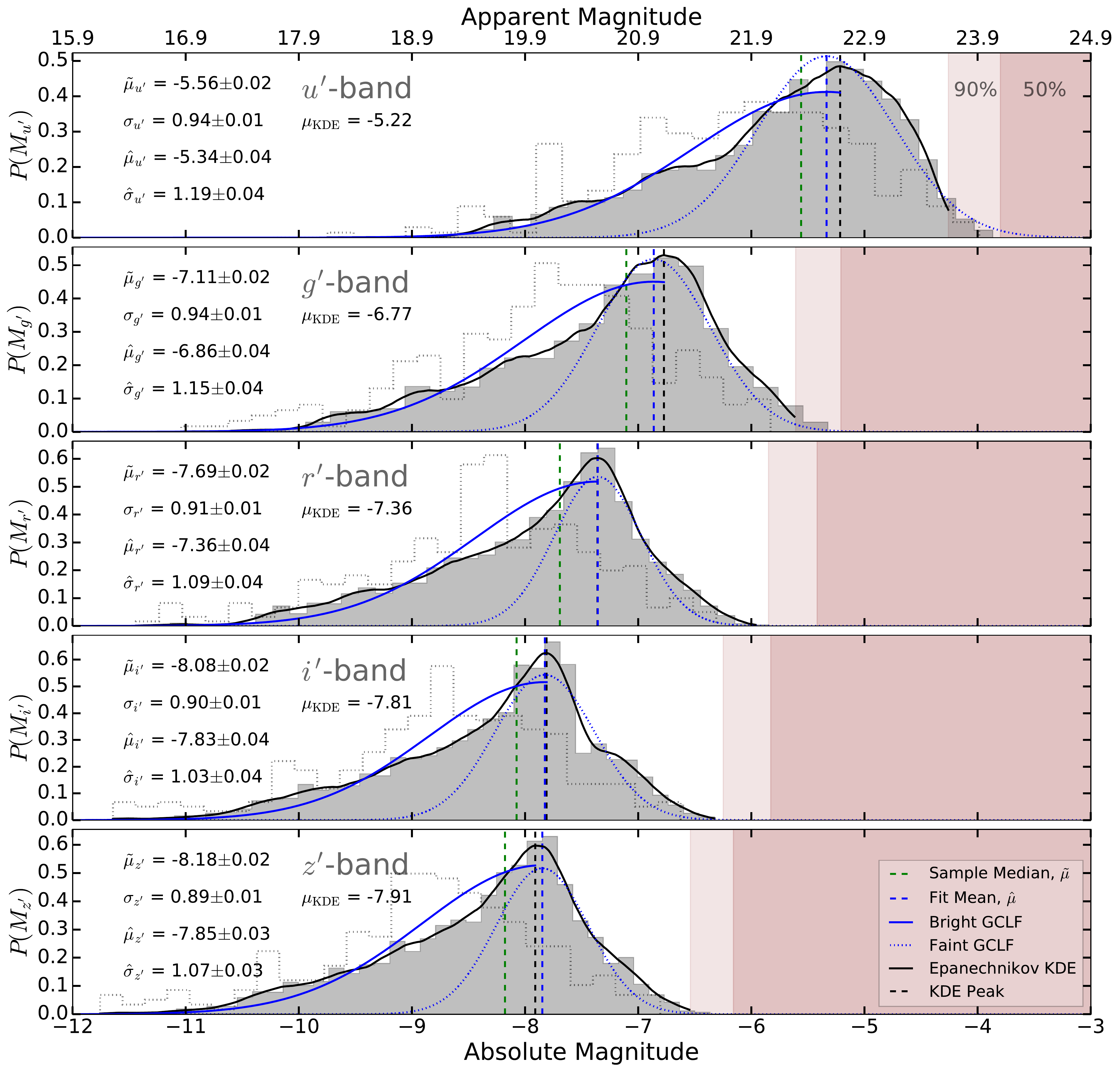}
\caption{Globular cluster candidate luminosity functions in the $u'$, $g'$, $r'$, $i'$, and $z'$ filters. Grey shading shows sample distributions using an optimal binning technique \citep{knu06}, and non-parametric Epanechnikov-kernel probability density estimates are shown by the solid black curves, limited to the 90\% completeness-limited data for the GCLFs (light red shading). Listed in the upper left corners are sample medians (dashed green lines) and standard deviations, alongside their $1\sigma$ bootstrapped error estimates. Alternatively, the solid blue curves show $N(\mu,\sigma)$ fits to the brightest/most massive candidates mirrored about the peak of the KDE (dashed black lines), with the corresponding $\hat\mu$ shown by the dashed blue lines and listed with the fit $\hat{\sigma}$. Similarly, the dotted blue curve shows a normal fit to the faintest candidates mirrored about the peak of the KDE, which does not represent the overall sample well. Lastly, we show by the dotted gray histograms the re-normalised distributions of previously confirmed GCs that failed to satisfy our selection algorithm, noting that these GCs are included in the solid shaded histograms. Point-source completeness levels corresponding to $\leq\!50\%$ is indicated by the darker red shading.}
\label{fig:gclfs}
\end{figure*} 

\section{Results}
\label{sec:results}
For the first time, a catalogue of confirmed GCs bolstered by a near-complete list of GC candidates reaching into the extreme halo of \cena\ is in-hand. Noting the homogeneous areal coverage of tiles\,1--7, attention now turns to the colour and spatial characteristics of the catalogue. The well established bimodal colour distributions of GC systems \citep[e.g.][]{sea78,zep93,ost93,whi95,els96,geb99,kun01,lar01,bro06,pen06,spi06,gou07,ric12} are often used to broadly infer the metallicities of the underlying GC stellar populations in that red GCs represent more metal-rich populations, while bluer GCs imply lower metallicities \citep[e.g.][]{puz05b}. In one paradigm, red GCs presumably form and are enriched during the initial giant starbursts that lead to the buildup of the majority of their host galaxy's stellar content. On the other hand, blue GCs preferentially form in more quiescent environments and/or shallower host potential wells \citep[i.e.\ dwarf halos, see][]{cot98,cot00,dab14a,dab14b,dab15}. However, this notion is under dispute, with lines of evidence suggesting that blue GCs form from pristine gas {\it in situ} with their giant hosts \citep[e.g.][]{ash92,for97b,bea02,spi06} with the red peak corresponding to later merger-induced starbursts.

With the above in mind, understanding GC luminosity and colour distributions, together with spatial variations, is useful to infer overall properties of their environment. The following sections discuss these distributions, beginning with colours/luminosities of the GC candidate catalogue (including the confirmed GCs and H12 survivors), followed by a spatial analysis. Finally, the two are combined in an effort to search for potential colour- and space-dependent patterns and/or features, as well as their implications on the assembly history of \cena\ and its surroundings.

%%%%%%%%%%%%%%%%%%%%%%%%%%%%%%%%%%%%%%%%%%%%%%%%%%
%%%%%%%%%%%%%%%%%%%%%%%%%%%%%%%%%%%%%%%%%%%%%%%%%%
%%%%%%%%%%%%%%%%%%%%%%%%%%%%%%%%%%%%%%%%%%%%%%%%%%

\subsection{Globular Cluster Luminosity Functions}
\label{sec:gclf}
The GC luminosity function (GCLF) describes the luminosity distribution of GCs, and is typically of log-normal form with a near-universal peak, or turnover, near $M_V\!\approx\!-7.4$\,mag \citep[${\cal M}_*\approx10^5\,M_\odot$; e.g.][]{jac92, ric95, har01, mcl05, bro06}. The five panels of Fig.\,\ref{fig:gclfs} illustrate the GCLFs for the $u'$, $g'$, $r'$, $i'$, and $z'$ filters. In all panels, the grey histograms show the distributions for GCs/candidates, which we estimate with non-parametric Epanechnikov-kernel probability density estimates\footnote{{\sc Python} package: {\sc scikit-learn}} (KDEs; solid black curves) to the 90 percent completeness limited data (light red shading). We indicate the KDE peaks by the dashed black lines, which we use to fit\footnote{{\sc Python} package: {\sc SciPy/optimize}} $N(\mu,\sigma)$ relations to the bright/high-mass and faint/low-mass candidates. First, we fit the bright GC candidates by mirroring them about the KDE peaks, and use the peaks and sample $\sigma$'s as initial guesses in the optimization. The solid blue curves show the resulting $N(\mu,\sigma)$ curves fit to the bright candidates, with the fit parameters ($\hat\mu$, $\hat\sigma$) listed below the sample values in each panel. The dashed green and blue lines indicate the sample medians, $\tilde\mu$, and means, $\hat\mu$, respectively. We then fit the faint samples by mirroring them about the KDE peaks, with the same initial guesses. The resulting dotted blue lines fail to represent the bright candidates well, and may be indicative of incompleteness toward fainter magnitudes, despite the 90 percent completeness estimates. This incompleteness may be a result of the final cut on {\sc spread\_model}, which may remove some faint, point-like GC candidates along with the foreground stars (see Fig.\,\ref{fig:galpoint}). In any case, this effect cannot be improved upon with the optical data without introducing significant stellar contamination to the catalogues, and will be discussed in greater detail in subsequent papers of this series, once our NIR data of the central SCABS field is fully analyzed. The combination of optical and NIR data will allow us to select GCs with the $uiK$ technique \citep[see][]{mun14} and reduce the contamination by foreground stars and background galaxies to a few percent.

We convert the $(g'-r')_0$ to the $V$-band using Eq.\,\ref{eq:grvconvert} and find that the $V$-band sample median of $\tilde\mu_0=-7.44\pm0.02$\,mag is $\sim0.36$\,mag brighter than the typical peak of $\sim-7.09$\,mag \citep[after correcting for $A_V=0.315$\,mag of foreground extinction;][]{sch11}, which may be affected by the high-luminosity tail shown by the KDE fits. Conversely, a $N(-7.09,1.13)$ fit to the bright GCs shows a mean $M_V$ that concurs with the peak KDE to the data, and recovers the expected $V$-band GCLF peak of $-7.09$\,mag. With that said, the non-parametric KDEs clearly follow the data more closely than the normal distributions, particularly in the case of the high-luminosity tails that are seen in each band.

The apparent bright tails might indicate that \cena\ is overabundant in massive GCs and/or UCDs, or that the candidate list still suffers some contamination from stellar sources. Given the dominance of $P({\rm GC})=1.0$ candidates in the final sample (see Fig.\,\ref{fig:urzprobs}, lower panel), and noting that the $P({\rm GC})\leq0.5$ candidates that make up $\lesssim6$ percent of the total catalogue have a range of luminosities (rather than being preferentially bright), supports the ruling out of the latter interpretation. At the same time, a reasonable concern regarding the former notion is that the failure to select $\sim60$ percent of the confirmed GC population might introduce a selection bias that could give rise to the high-luminosity tail. To check against this effect, the grey dotted histograms in Fig.\,\ref{fig:gclfs} show the $\sim300$ bona-fide GCs that did not survive the selection algorithm, but were nonetheless recovered in our imaging. It is important to note that this sample is already represented in the grey shaded distributions, and the dotted histograms are re-normalized to better compare to the overall sample. In each filter, the unrecovered, but bona-fide GCs show the high-luminosity tail, suggesting that the overall sample statistically resembles the bright end of \cena's GCLF, and that it is likely to be overabundant in massive GCs/UCDs. In any case, further refinement of the GC candidate catalogue will await either large-scale spectroscopic campaigns and/or the extension of the photometric SED coverage.

\begin{figure}
\centering
\includegraphics[width=\linewidth]{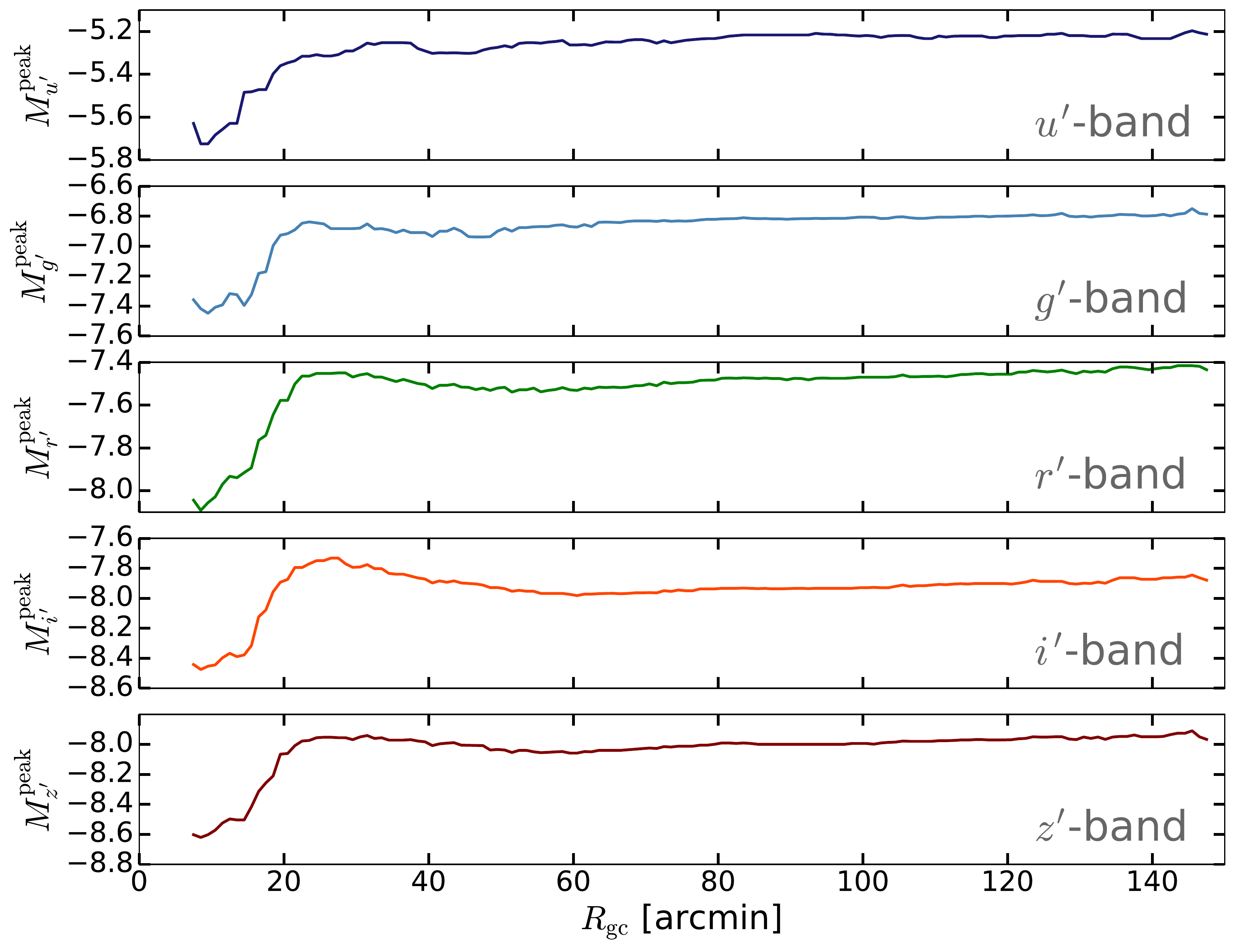}
\caption{Evolution of the GCLF peaks with galactocentric radius. The peak magnitudes of the $u'g'r'i'z'$ GCLFs are derived from KDE fits to GCs binned in rolling $R_{\rm gc}\pm5\arcmin$ probing \cena's halo in $1\arcmin$ steps.}
\label{fig:rgc_gclf}
\end{figure} 

Dynamical friction arguments imply that the GCLF turnover for dwarf galaxies should become slightly fainter over time with respect to giants \citep[e.g.][]{lotz01,bek10}. Observational evidence suggests that dwarf galaxy GCLF peaks can be as much as $M_V\sim0.4\,{\rm mag}$ fainter than the giant $M_V\approx-7.4\,{\rm mag}$ turnover \citep[e.g.][]{dur96,mil07,geo09,vil10}; however, it must be acknowledged that there also exists evidence to the contrary, where dwarf GCLF peaks--at least for dwarf irregulars--are indistinguishable from giants \citep[e.g.][]{set04,str06,geo09}. Fig.\,\ref{fig:rgc_gclf} investigates this idea by looking at the behaviour of the \cena\ GCLF shape as a function of $R_{\rm gc}$. The GC candidates are binned in rolling windows of $\Delta R_{\rm}=10\arcmin$ in steps of $1\arcmin$, and the peak of a KDE fit is determined for each bin individually. The peaks are shown for all five GCLFs against $R_{\rm gc}$ where a noticeable $\sim\!0.5\!-\!0.6\,{\rm mag}$ jump is seen for the $g'r'i'z'$ filters in the region $15\arcmin\lesssim R_{\rm gc} \lesssim 30\arcmin$, which becomes less pronounced toward the bluer bands. Beyond $R_{\rm gc}\approx30\arcmin$, each declines by $\sim\!0.2\!-\!0.3\,{\rm mag}$, before gently, and almost monotonically, becoming fainter again by $\sim\!0.2\!-\!0.3\,{\rm mag}$. A similar pattern is seen in the $u'$-band GCLF, but with a less dramatic $\Delta M_{u'}\approx0.3-0.4\,{\rm mag}$ change toward the outer regions accompanying increased scatter likely due to a combination of larger photometric uncertainties, heightened sensitivity to younger/metal-poor stellar populations, and/or the existence of HHBs. Altogether, the diminishing GCLF peak luminosity toward the outer halo of \cena\ is broadly consistent with GC origins in lower-mass dwarf halos, the implications of which are discussed in \S\,\ref{sec:gctotal}, but we defer a more detailed look at the $15\arcmin\lesssim R_{\rm gc}\lesssim 30\arcmin$ region to a future work.

\begin{figure*}
\centering
\includegraphics[width=\linewidth]{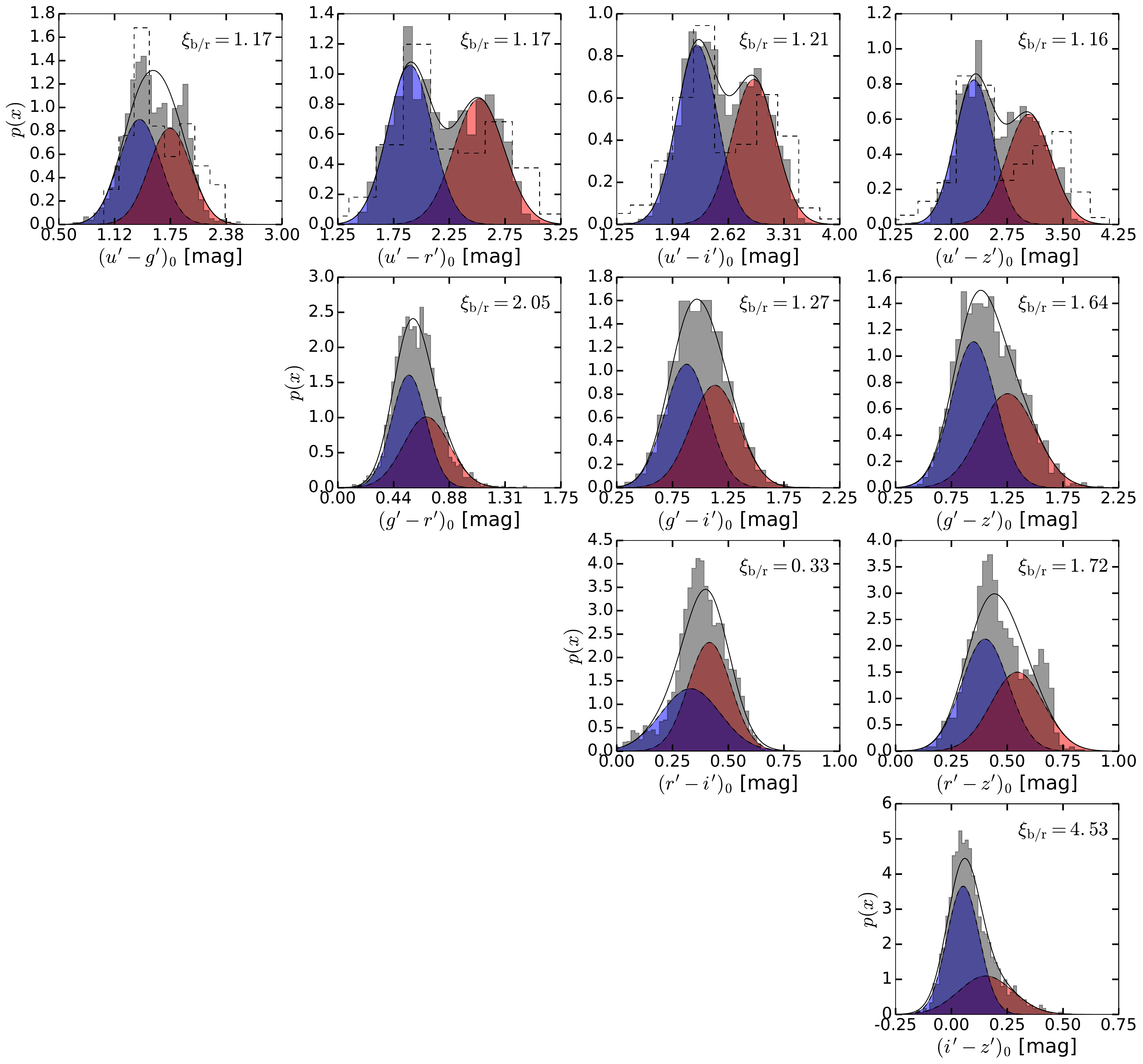}
\caption{Globular cluster colour distributions. Optimally binned GC candidate distributions are shown by the grey shading for each possible colour combination with filters running through $u'$, $g'$, $r'$, $i'$, and $z'$ from top to bottom, and left to right. Over-plotted are best-fit GMMs, which sum to the black curves shown. The ratios of blue GC candidates to red are indicated on each panel, which are based on the automatic classification from the respective GMMs. e shows a normal fit to the faintest candidates mirrored about the peak of the KDE, which does not represent the overall sample well. The dashed gray histograms in the upper panels show the $u'$-based colour distributions of previously confirmed GCs that failed to satisfy our selection algorithm, noting that these GCs are included in the solid shaded histograms. Final blue/red classification is based on the $(u'-z')_0$ colour.}
\label{fig:col_distros}
\end{figure*} 

\subsection{Globular Cluster Colour Distributions}
\label{sec:colour_distros}
Distributions of each colour permutation are shown in Fig.\,\ref{fig:col_distros} with bluer filters corresponding to upper rows, and colour indices probing wider SED coverage from left to right. Clear bi-modalities are visible from the binned data (grey histograms), which are prominent when the NUV $u'$ filter is used (top row), and become less significant with smaller SED sampling width. Single, double, and triple component Gaussian mixture models (GMMs) are fit\footnote{{\sc Python} package: {\sc scikit-learn}} to each colour combination, with the best model chosen based on the minimization of the Bayesian Information Criterion, which is a penalized method of model selection \citep{sch78}. In all cases a two-component GMM is preferred (solid black lines), which is decomposed into the dashed black lines and correspondingly coloured shading. The GMMs are used to classify GCs as either blue or red, and $(r'-i')_0$ is the sole index where the number ratio of blue to red GCs, \btrfrac, is below unity. It is clear that NUV photometry strongly aids the separation of blue and red GCs, as shown by the consistent \btrfrac$=1.16-1.21$ when using NUV colour indices, compared to the widely varying \btrfrac\ predicted by non-NUV based colours. We thus exploit the wide SED sampling of the $(u'-z')_0$ colour index to classify the GC catalogue and adopt \btrfrac$=1.16$ for the total sample.

We note that, like the GCLFs, the significant culling of bona-fide GCs via the strict selection criteria risks introducing biases into the overall colour distributions. To investigate this, we show by the dashed histograms in the upper panels of Fig.\,\ref{fig:col_distros} the $u'$-based colour distributions for those confirmed GCs that did not survive the selection algorithm. By comparison to the overall samples, it is clear that the bi-modality and relative blue/red peak heights are broadly consistent, indicating that it is unlikely that the selection criteria introduces strong biases in the colour distributions and classification of the new GCs as blue or red.

It is well established that the mean metallicity of GC systems increases with the metallicity of their hosts, with correspondingly redder mean colours accompanying higher host mass/luminosity/metallicity \citep[e.g.][]{van75,bro91,for97b,lar01,bur01,lot04,pen06}. At the same time, \cite{pen06} showed that \btrfrac\ for gE galaxies in the Virgo cluster decreases with luminosity for galaxies in the magnitude range $-22<M_B<-15$\,mag. Given \cena's $M_B=-20.1$\,mag \citep{lau89}, then if it had a similar evolution as Virgo giants, one could expect \btrfrac\ slightly less than what is presently observed. With that said, a smaller \btrfrac$\approx1.11$ has already been measured for a sample of 194 \cena\ GCs by \cite{har04b}; however, their results were based on metallicity sensitive Washington photometry, and were limited to $R_{\rm gc}\!=\!45\arcmin$, so a direct comparison to the present results might not be warranted. Even so, limiting our sample to only those GC candidates within $45\arcmin$ results in \btrfrac$=1.30$.

Given that characteristics of metal-rich GC populations correlate with the overall host properties, while the metal-poor component has fewer such dependencies, then the relative strength of the blue population compared to red likely depends on the merger and star-formation history of the host \citep[][]{for97a,for97b,cot98,mas10,arn11,ton13} and will most likely be also a function of galactocentric radius. The high \btrfrac\ shown by \cena\ then provides a strong clue to the dominant mechanism behind its mass assembly. Table\,\ref{tbl:gccolours} lists some simple statistics of the blue and red components of \cena's GC sample, with information shown for the population as a whole, and separated based on proximity to the host. For comparison with the results of \cite{pen06}, the median and mean $(g'-z')_0$ colours are listed for each component and we find a median blue $(g'-z')_0$ colour of $\tilde{\mu}_{\rm blue}\approx0.90$\,mag to be consistent with the expectation for a gE of \cena's luminosity in Virgo. The same cannot be said of the red GCs, in that at all radii $\tilde{\mu}_{\rm red}\approx1.26$ is $\sim0.02-0.06$\,mag bluer and more consistent with a Virgo gE fainter than \cena. Taken together, this result might be suggestive of the blue population having been assembled through long-term accretion processes and/or tidal stripping of dwarfs, whereas the red population was assembled by a major merger of galaxies both similar in mass. 

Further study of the GC colour distributions is deferred to \S~\ref{sec:gctotal} where it is combined with the results of the spatial distribution analysis described in the following section.

\begin{table}
	\centering
	\caption{Colour statistics for the GC catalogue. Col.\,(1) lists the sample being considered, followed by median and mean $(g'-z')_0$ colours for the blue and red components, and the ratio of blue GCs to red based on the $(u'-z')_0$ classification. All parameters have 0.01\,mag $1\sigma$ bootstrapped errors.}
	\label{tbl:gccolours}
	\begin{tabular}{lccccc}
		\hline\hline
		Sample 	& 	$\tilde{\mu}_{\rm blue}$	&	$\tilde{\mu}_{\rm red}$	&	$\mu_{\rm blue}$	&	$\mu_{\rm red}$	&	\btrfrac	\\
				&	(mag)				&	(mag)				&	(mag)			&	(mag)			&					\vspace{1mm}\\
		\hline
		Total					&	0.90		&	1.26		&	0.92		&	1.28		&	1.16		\\
		$R_{\rm gc} < 50'$		&	0.88		&	1.27		&	0.88		&	1.27		&	1.33		\\
		$R_{\rm gc} \geq 50'$	&	0.91		&	1.26		&	0.94		&	1.28		&	1.08		\\
		\hline\hline
	\end{tabular}
\end{table}

%%%%%%%%%%%%%%%%%%%%%%%%%%%%%%%%%%%%%%%%%%%%%%%%%%
%%%%%%%%%%%%%%%%%%%%%%%%%%%%%%%%%%%%%%%%%%%%%%%%%%
%%%%%%%%%%%%%%%%%%%%%%%%%%%%%%%%%%%%%%%%%%%%%%%%%%
\begin{table}
	\centering
	\caption{Results of the linear regression analysis for the GC catalogue. Col.\,(1) lists the sample under consideration, while cols.\,(2)--(7) show the best fit power slope and the associated variance score for the total sample within $R_{\rm gc}\leq120\arcmin$, the inner ($R_{\rm gc}\lesssim50\arcmin$) sample, and the outer ($70\arcmin\lesssim R_{\rm gc}\lesssim 120\arcmin$) candidates.}
	\label{tbl:gcspace}
	\begin{tabular}{lcccccc}
		\hline\hline
		Sample 	& 	$\Gamma_{\rm all}$	&	$r^2_{\rm all}$	&	$\Gamma_{\rm inner}$	&	$r^2_{\rm inner}$	&	$\Gamma_{\rm outer}$	&	$r^2_{\rm outer}$	\vspace{1mm}\\
		\hline
		Total		&	$-$1.22	&	0.95		&	$-$1.55	&	0.99		&	$-$0.48	&	0.29		\\
		Blue		&	$-$1.25	&	0.65		&	$-$1.40	&	0.99		&	$-$0.61	&	0.43		\\
		Red		&	$-$1.19	&	0.40		&	$-$1.78	&	0.98		&	$-$0.33	&	0.11		\\
		\hline\hline
	\end{tabular}
\end{table}

\subsection{Globular Cluster Spatial Distributions}
\label{sec:spatial_distros}
\begin{figure*}
\centering
\includegraphics[width=\linewidth]{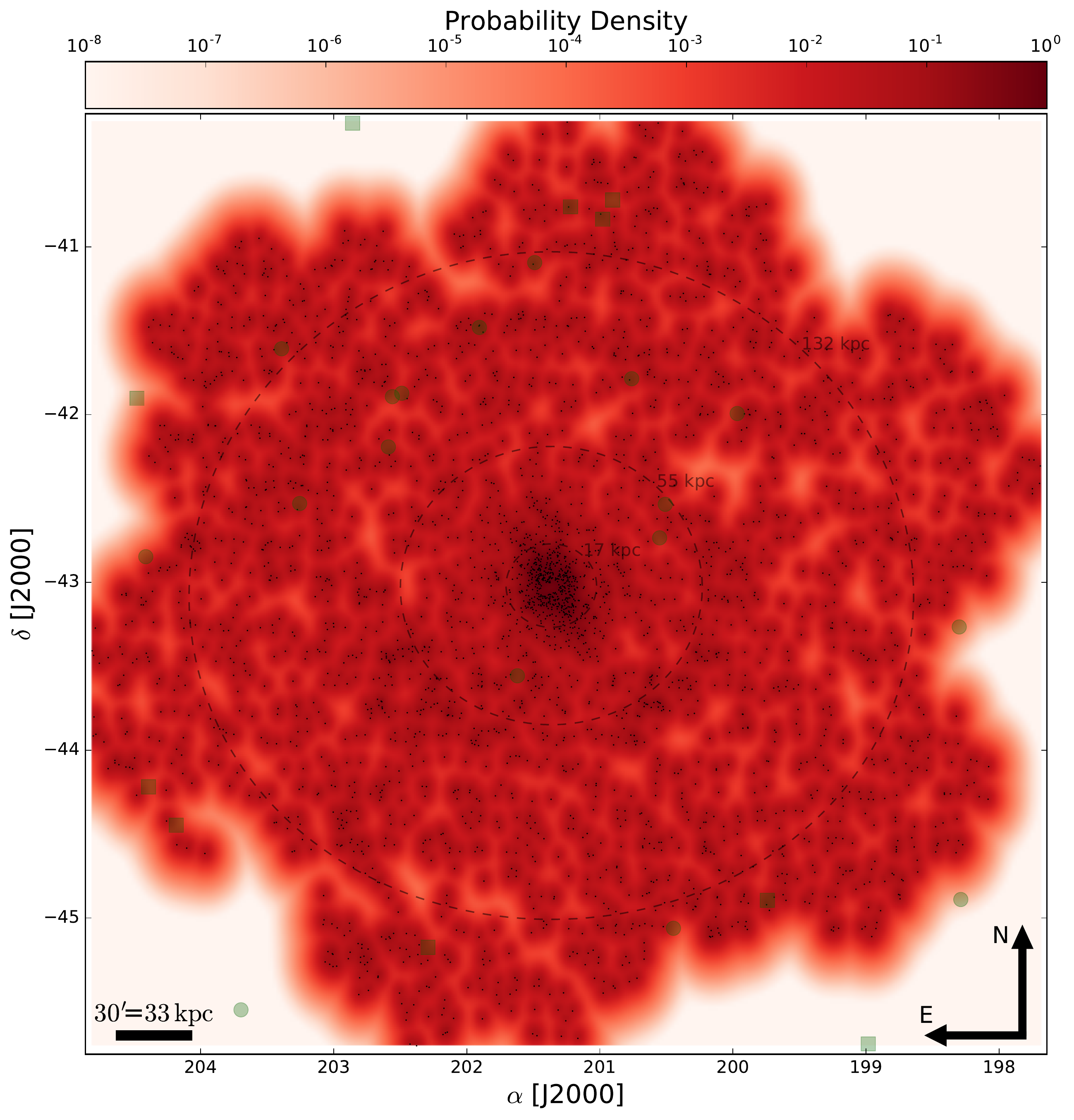}
\caption{The spatial distribution of all globular cluster candidates. The coordinates of all GC candidates are shown by the small black points, which strongly cluster around \cena's location in the centre of the figure. The colour map represents a 2D exponential-kernel probability density estimate to the data, where the darker shading indicates regions of projected GC spatial over-densities. Over-plotted on the figure is the confirmed dwarf galaxy population \citep[dark green dots; ][]{cot97,van00,kar07,crn14,crn15}, and newly identified dwarf candidates \citep[green squares; ][]{mul15,mul17}. To guide the eye, ellipses are drawn at $R_{gc}=15\arcmin$, $50\arcmin$, and $120\arcmin$ and labelled with the corresponding physical scales. We note the sharp decline in the probability density near the edge of the tile\,1--7 fields-of-view.}
\label{fig:gcdense}
\end{figure*} 

An alternative but equally important property of a GC system is its spatial distribution. An intriguing trend of giant galaxy GC systems is that redder GCs tend to cluster toward smaller $R_{\rm gc}$ than their bluer counterparts that may have been deposited at larger radii during satellite accretion events \citep[e.g.][]{gei96,for97b,ash98,cot98,cot00,for01,puz04,bas06,bro06,spi06,gou07,fai11,for12,dab14a,dab14b,dab15,kar16}.

As a first look at the global distribution of GCs and candidates around \cena, Fig.\,\ref{fig:gcdense} shows a scatter plot of the GC catalogue (black dots), with the colour indicating a non-parametric exponential-KDE\footnote{{\sc Python} package: {\sc scikit-learn}} with a $0.025^\circ$ bandwidth. While the distribution is generally unstructured, the GC distribution closest to \cena\ appears slightly elongated along the major isophotal axis along azimuthal angles $\Phi=35^\circ/215^\circ$ (E leading N) of the host, as noted in previous works \citep[e.g.][]{har04b,pen04b,woo07,woo10a}. Recognizing the absence of spatial bias in the present data, this effect appears to be a real feature of \cena's GC system; however, the bias against the minor axis disappears outside of $\sim15-20\arcmin$. 

Beyond $R_{\rm gc}\approx15\arcmin$ ($\sim17$\,kpc), Fig.\,\ref{fig:gcdense} shows a faint, broad over-density centred at $(\alpha,\delta)\!\approx\!(202.5^\circ,-43.5^\circ)$, extending in a counterclockwise arc to coordinates $(\alpha,\delta)\!\approx\!(200.0^\circ,-43.0^\circ)$. An equivalent broad over density is not seen between $R_{\rm gc}\approx30-80\,{\rm kpc}$ NE of \cena, but a tenuous connection to another $\Delta R_{\rm gc}\approx 10\,{\rm kpc}$-thick arc can be seen spiraling clockwise from $(\alpha,\delta)\approx(203.5^\circ,-42.5^\circ)$ to $(\alpha,\delta)\approx(200.5^\circ,-41.7^\circ)$, N of the host. Interestingly, the NE arc projects directly across at least two known dwarfs (green dots) in the system, and could be associated with a disrupting dwarf in the system \citep[][their Dw3]{crn15}, although the lack of 3D information precludes drawing conclusions on possible physical associations. While intriguing, a more detailed analysis of the significance of these features is deferred to a future work.

\begin{figure*}
\centering
\includegraphics[width=0.9\linewidth]{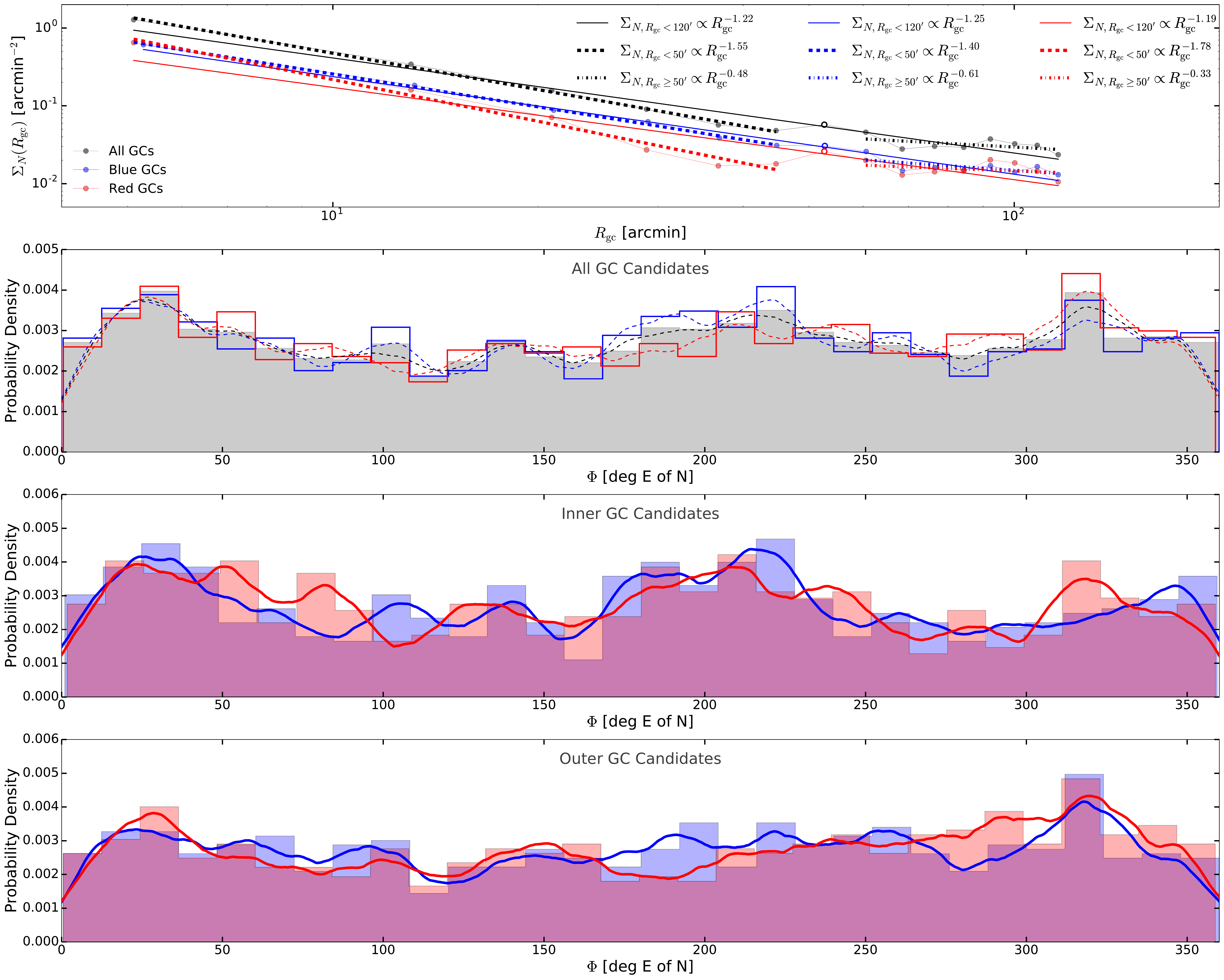}
\caption{Radial and azimuthal distributions of GC candidates. ({\it Top)}: The $\Sigma_N(R_{\rm gc})$ profile calculated along radial bins, with solid lines indicating power-law fits to all GCs with $R_{\rm gc}\leq120'$, and dashed and dot-dashed curves showing fits to data inside and outside of $\sim55\arcmin$. Black curves represent all GC candidates, while blue and red are coded to the respective GC subsamples in all panels. In all fits we exclude the anomalously high $\Sigma_N(R_{\rm gc})$ points centred at $R_{\rm gc}\simeq 55\arcmin$ (open circles) to avoid a potential bias arising from the over density discussed in the text. ({\it Second from top}): Distribution of $\Phi$ in units of degrees E from N with the grey shading representing the total population. Solid lines outline the binned data, while Epanechnikov-KDEs are shown by the dashed curves. ({\it Second from bottom}):\,Same as the previous panel, but with blue/red histograms indicating the blue/red samples with $R_{\rm GC}<50\arcmin$. The coloured histograms are shown with lowered opacities so that regions of overlap (darker shading) are highlighted, and solid curves show Epanechnikov-KDEs fit to the data. ({\it Bottom}): Same as the previous panel, but showing only the blue/red samples with $R_{\rm GC}\geq50\arcmin$.}
\label{fig:surfazi}
\end{figure*} 

\subsubsection{Radial Surface Number Density Profile}
Fig.\,\ref{fig:surfazi} shows alternative visualizations of the spatial distributions of the GC candidates, considering only those within $R_{\rm gc}\leq120\arcmin$ to avoid artificial biases arising from inhomogeneities at the edge of the SCABS tiles (1-7, see Figs.~\ref{fig:dither} and \ref{fig:gcdense}). The top panel shows the projected radial surface number density profile, $\Sigma_N(R_{\rm gc})$, calculated by individually binning the total, blue, and red subsamples, and determining
\begin{equation}
\Sigma_N(R_{\rm gc})=\frac{N_{\rm GC}}{\pi\left(R_{{\rm gc},o}^2-R_{{\rm gc},in}^2\right)}
\end{equation} 
in radial annuli, where $R_{{\rm gc},o}$ and $R_{{\rm gc},in}$ are the outer and inner radii of each annulus. A linear regression analysis\footnote{{\sc Python} package: {\sc scikit-learn}} is applied to fit a linear relation in logarithmic space to determine the power law shapes that best explain the data, with results listed in Table\,\ref{tbl:gcspace}. We note that in what follows we exclude the bins centred at $R_{\rm gc}\simeq 55\arcmin$ to avoid artificially biasing the results with the ring-type over-density mentioned above, which is represented in the top panel of Fig.\,\ref{fig:surfazi} by the notable uptick shown by the open circles. First, a single power-law of the form,
\begin{equation}
\Sigma_N(R_{\rm gc})\propto R_{\rm gc}^\Gamma
\end{equation}
is fit to all of the data (solid curves) whose $r^2$ scores indicate that a power-law slope of $\Gamma=-1.22$ explains 95 percent of the total sample variance, while $\Gamma=-1.25$ and $-1.19$ only explain $65$ percent and $40$ percent of the blue and red variances, respectively.

If the entire radial sample is considered, the three $\Sigma_N(R_{\rm gc})$ profiles show power-laws with similar slopes. Apart from the sample as a whole, these relations do not appear to adequately explain the data. Splitting the GC candidate sample into ``inner'' ($R_{\rm gc}<50\arcmin$) and ``outer'' ($60\arcmin\leq R_{\rm gc}\leq120\arcmin$) populations, and applying the linear regression analysis to each individually shows a different picture. Here the fits explain the data better, finding that $\Gamma=-1.55$, $-1.40$, and $-1.78$ accounts for $\gtrsim98$ percent of the total, blue, and red variances, respectively, for the inner sample. Meanwhile, the fits for the outer subsamples all show shallower $\Gamma=-0.48$, $-0.61$, and $-0.33$ slopes, but with much more penalized $r^2$ scores that are likely to arise from the substructure noted above.

In accords with previous works on \cena's and other GC systems, we find that the inner red population shows a steeper relations than the blue. Conversely, the outer samples show the opposite behaviour. All profiles flatten, but more sharply for the red GCs, its poor $r^2=0.11$ score notwithstanding. The proximity of \cena\ makes it among only a few gEs for which resolved stellar populations are possible. In particular, recent works have studied the stellar distributions out to similar radii as this work, finding red giant branch stars to populate the halo out to at least 140\,kpc \citep[e.g.][]{crn13,rej14}. Unfortunately, a robust quantitative comparison between GC and stellar radial surface number density profiles is complicated by differing tidal stripping timescales between GCs and galaxy spheroid stars during merger events \citep[e.g.][]{smi13} combined with the--possibly spatially biased--likely underestimate of the total GC population arising from our conservative selection procedure. Nonetheless, we point out qualitatively that the flattening of $\Sigma_N$ beyond $R_{\rm gc}\approx50-60\arcmin$ compares nicely to a flattening of the stellar profile beyond this radius \citep[][]{crn13}. Meanwhile, the difficulty in fitting a smooth power-law to GCs in the outer halo intriguingly hints at the strong spatial variability of RGB metallicities out to $R_{\rm gc}\approx140\arcmin$ \citep[][]{rej14}, especially if such features arise from disturbed substructure from prior merger events. Regardless, a robust statistical comparison between GC and stellar spatial/metallicity distributions in \cena's extreme halo deserves its own dedicated effort, and so we must defer it to a future work.

\subsubsection{Azimuthal Distributions}
The bottom three panels of Fig.\,\ref{fig:surfazi} show distributions of azimuthal angle, $\Phi$, in degrees East of North. The second panel from the top shows the distributions of the total population (grey shading), with the blue and red samples over-plotted by respectively coloured solid lines. Epanechnikov-KDEs with $15^\circ$ bandwidths are shown by the dashed lines. The ``inner'' and ``outer'' (based on a $R_{\rm gc}=50\arcmin$ cut) populations are shown separately in the bottom two panels respectively, where the blue and red populations are indicated by the respective colours with opacity increased to show regions of overlap. The data are again smoothed by $15^\circ$ KDEs, which are represented by the solid blue and red relations.

The overall population shows a mildly bimodal structure corresponding to the major axis along the $\Phi\!\approx\!35^\circ$--$\Phi\!\approx\!215^\circ$ line, with indications for another peak near $\Phi\approx315^\circ$. Considering the inner and outer samples separately shows that the bimodal structure is more prevalent for the inner GCs, with only a mild over density seen near $\Phi\!\approx\!35^\circ$ for both red and blue samples, and only the blue GCs showing any indication for an elevated surface number density near $\Phi\approx200-215^\circ$. While the GC system seems to only show mild evidence for bi-modality along \cena's major axis out to $R_{\rm gc}\gtrsim120\arcmin$, it does appear consistent with the ellipticity already reported for its resolved red giant star population at similar scales \citep[][]{crn13,rej14}. One other feature of note is that seen at $\Phi\approx315^\circ$, which is notable in the inner sample for only the red subsample, and is more significant in the outer sample, where both blue and red candidates show signs of an over density. In general, despite the steeper radial surface number density profile shown by the red GCs for the inner sample, the azimuthal distributions show relatively similar behaviours at all radii, with only mild indications for azimuthal sub-peaks that would indicate clustering on different length scales.

\begin{figure}
\centering
\includegraphics[width=\columnwidth]{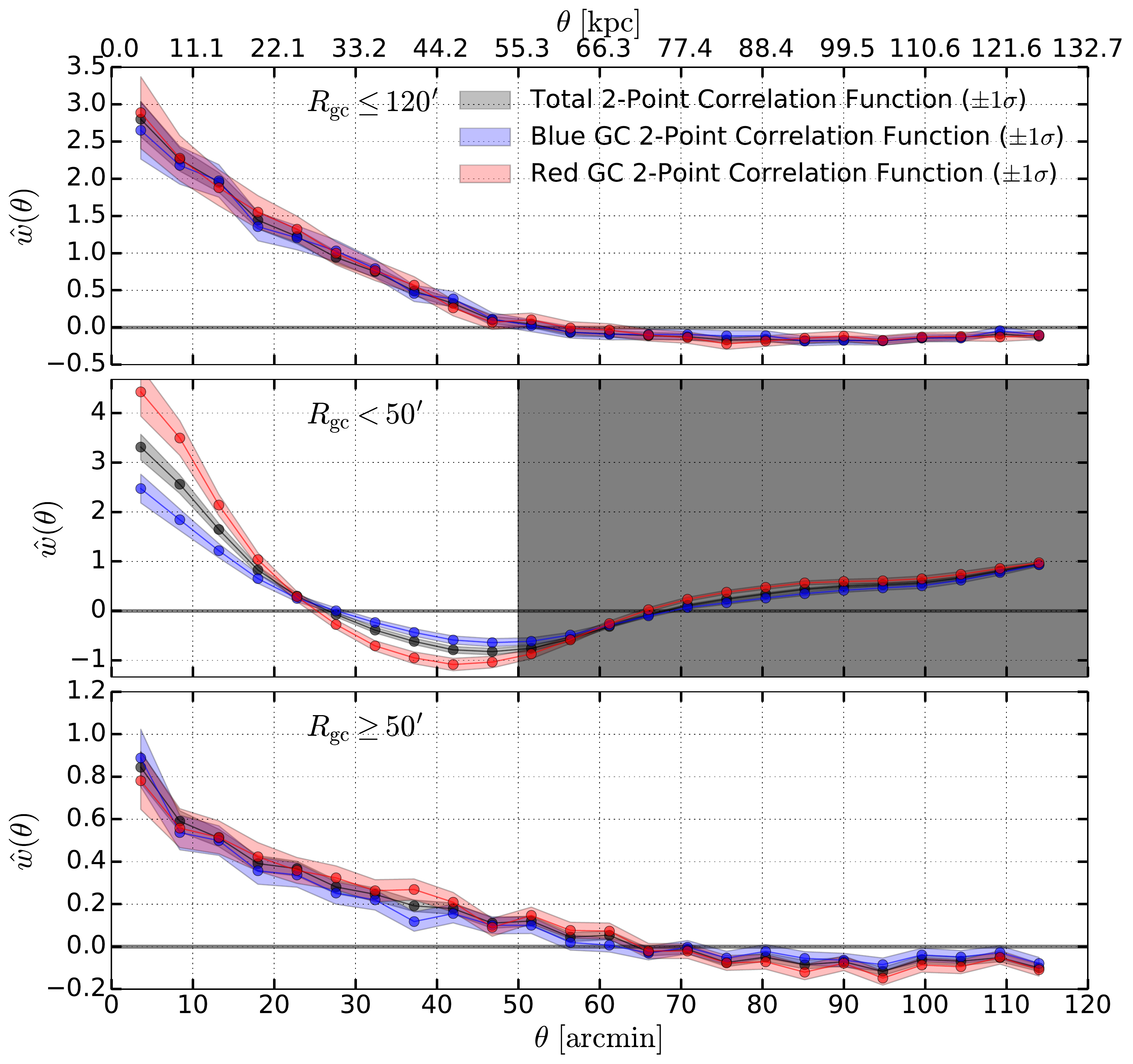}
\caption{Two-point angular correlation function analysis of the GC candidates. ({\it Top}):\,Two-point correlation functions for the total, blue, and red populations are indicated by the corresponding curves, with shaded $\pm1\sigma$ bootstrapped errors. ({\it Middle}):\,Same as above, but with clustering only considered for the GC samples within $R_{\rm gc}<50\arcmin$ of \cena. Note that only clustering on scales of $<50\arcmin$ is considered. ({\it Bottom}):\,The analysis applied to the outer ($R_{\rm gc}\geq50\arcmin$) subsamples. In all panels, the lower abscissa indicates $R_{\rm gc}$ in angular units, while the upper shows physical projected distance.}
\label{fig:2pt}
\end{figure} 

\subsubsection{Two-Point Correlation Functions}
The central clustering and potential arc(s) of GCs around \cena\ are the most obvious coherent GC structures in Fig.\,\ref{fig:gcdense}, and here we attempt to quantify potential levels of clustering. To this effect, the results of a two-point angular correlation function \citep[$w(\theta)$;][]{lan93} analysis\footnote{{\sc Python} package: {\sc astroML}} are shown in Fig.\,\ref{fig:2pt}. Briefly, given a set of points on the sky, $w(\theta)$ describes the level and scale of clustering by considering the probability of finding two points separated by an angle $\theta$ compared to what would be expected from a random distribution (i.e.\ $w(\theta)\approx0$). It should be stressed that in this case, $\theta$ represents the angular distance from each point under consideration, and not from the centre of \cena, i.e.\ $R_{\rm gc}=0\arcmin$.

The top panel of Fig.\,\ref{fig:2pt} shows $w(\theta)$ calculated for the total, blue, and red populations. For all samples, there exists a particularly strong probability of clustering on scales $\lesssim20$\,kpc, which is likely dominated by the concentration of GCs directly around \cena. Interestingly, all three samples show very similar behaviour, with evidence for clustering on all scales $\lesssim40$\,kpc, above which there is a marginal likelihood of clustering. This result can be visually verified by Fig.\,\ref{fig:gcdense}, where there is concentrated clustering in the centre of the field, with multiple smaller clumps throughout the region, and a few larger structures that often correspond to sections of the ``arc''.

As implied by Fig.\,\ref{fig:surfazi}, if the ``inner'' and ``outer'' GC populations are of different natures, then they might show different indications of clustering. To investigate this, the middle and bottom panels of Fig.\,\ref{fig:2pt} show $w(\theta)$ corresponding to the ``inner'' and ``outer'' populations, respectively. The middle panel shows that for the $R_{\rm gc}<50\arcmin$ GCs, the strong clustering of the core population dominates $w(\theta)$ such that clustering is most significant on scales $\lesssim20$\,kpc. Moreover, the inner, red population shows a higher probability of clustering on small scales, consistent with their steeper $\Sigma_N(R_{\rm gc})$ profile. Conversely, concurrent with the shallower blue GC slope, there is less evidence for clustering at the smallest scales, with $w_{\rm blue}(\theta)$ declining less sharply than the red sample out to $R_{\rm gc}\approx20$\,kpc, which represents the extent of \cena's inner halo. As a side-note, while outside of $R_{\rm gc}\approx20$\,kpc there is no evidence whatsoever for GC clustering, all samples show $w(\theta)\gtrsim0$ outside of $\approx60\arcmin$. This feature is a censoring artifact, and is to be expected since outside of the largest scale of interest (i.e.\ $R_{\rm gc}\gtrsim50\arcmin$; dark grey shading), there exists a growing probability of finding points separated by $\lesssim50\arcmin$ compared to what is expected of a uniform distribution within the larger region where there are artificially no points.

The bottom panel of Fig.\,\ref{fig:2pt} paints a somewhat different picture. Here the ``outer'' GCs show significant clustering on scales $\lesssim50\arcmin$. This result can be seen visually on Fig.\,\ref{fig:gcdense}, where the KDE shows higher density regions in the outskirts of the field covering a wide range of spatial scales. Interestingly, in this case, it is the blue population that hints at a slightly higher probability of clustering at the smallest scales together with a steeper decline. This feature is consistent with the steeper outer $\Sigma_N(R_{\rm gc})$ profile shown in Fig.\,\ref{fig:surfazi} (top panel). Altogether, the evidence of sub-$10$\,kpc scale clustering of blue GCs supports the model in which they primarily come from dwarf galaxies and may hint at a large reservoir of undiscovered or previously disrupted dwarf satellites in the extreme halo of \cena, similar to what is predicted by the $\Lambda$\,Cold Dark Matter cosmological framework \citep[][]{kly99,moo99} and hinted at in recent surveys of nearby galaxy clusters \citep{mun15,san16}.

%%%%%%%%%%%%%%%%%%%%%%%%%%%%%%%%%%%%%%%%%%%%%%%%%%
%%%%%%%%%%%%%%%%%%%%%%%%%%%%%%%%%%%%%%%%%%%%%%%%%%
%%%%%%%%%%%%%%%%%%%%%%%%%%%%%%%%%%%%%%%%%%%%%%%%%%

\section{Discussion}
\label{sec:discussion}
\subsection{Characteristics of the inner and outer \cena\ GC System}
\label{sec:colourspace}
\begin{figure}
\centering
\includegraphics[width=\columnwidth]{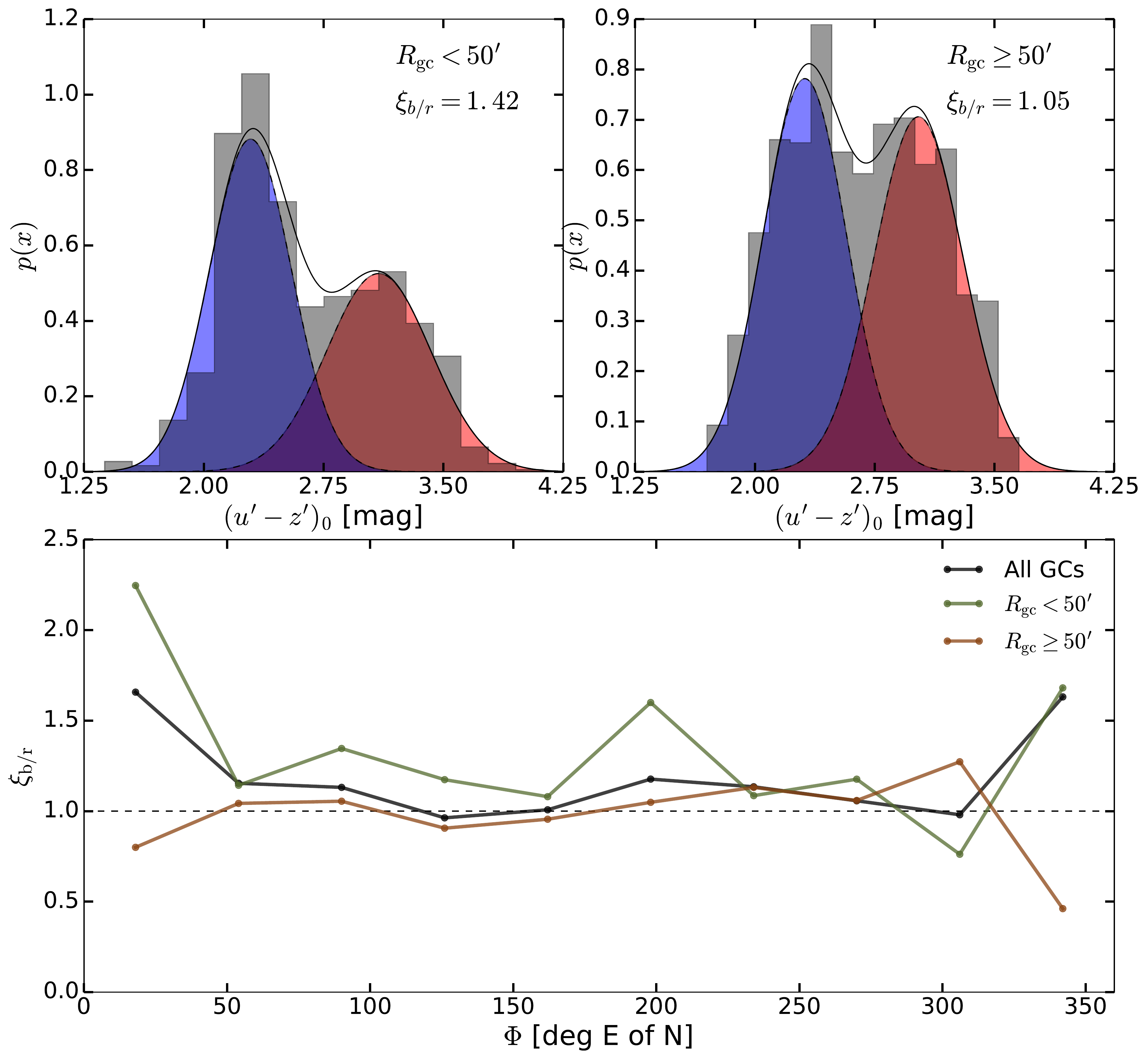}
\caption{The inner and outer blue-to-red GC candidate fractions. ({\it Upper panels}):\,Two-component GMM fits (solid black curves) to the inner ($R_{\rm gc}<50\arcmin$; left panel) GC candidates, and the outer ($R_{\rm gc}\geq50\arcmin$; right), with grey shading representing the underlying data. Blue/red shading represents the corresponding blue and red gaussian components. ({\it Lower panel}):\,\btrfrac\ is shown as functions of $\Phi$ in $36^\circ$-wide bins. The total population within $120\arcmin$ of \cena\ is shown by the black relation, while the green and brown relations show results for the inner and outer samples, respectively.}
\label{fig:inout_frac}
\end{figure} 

\begin{figure*}
\centering
\includegraphics[width=0.9\linewidth]{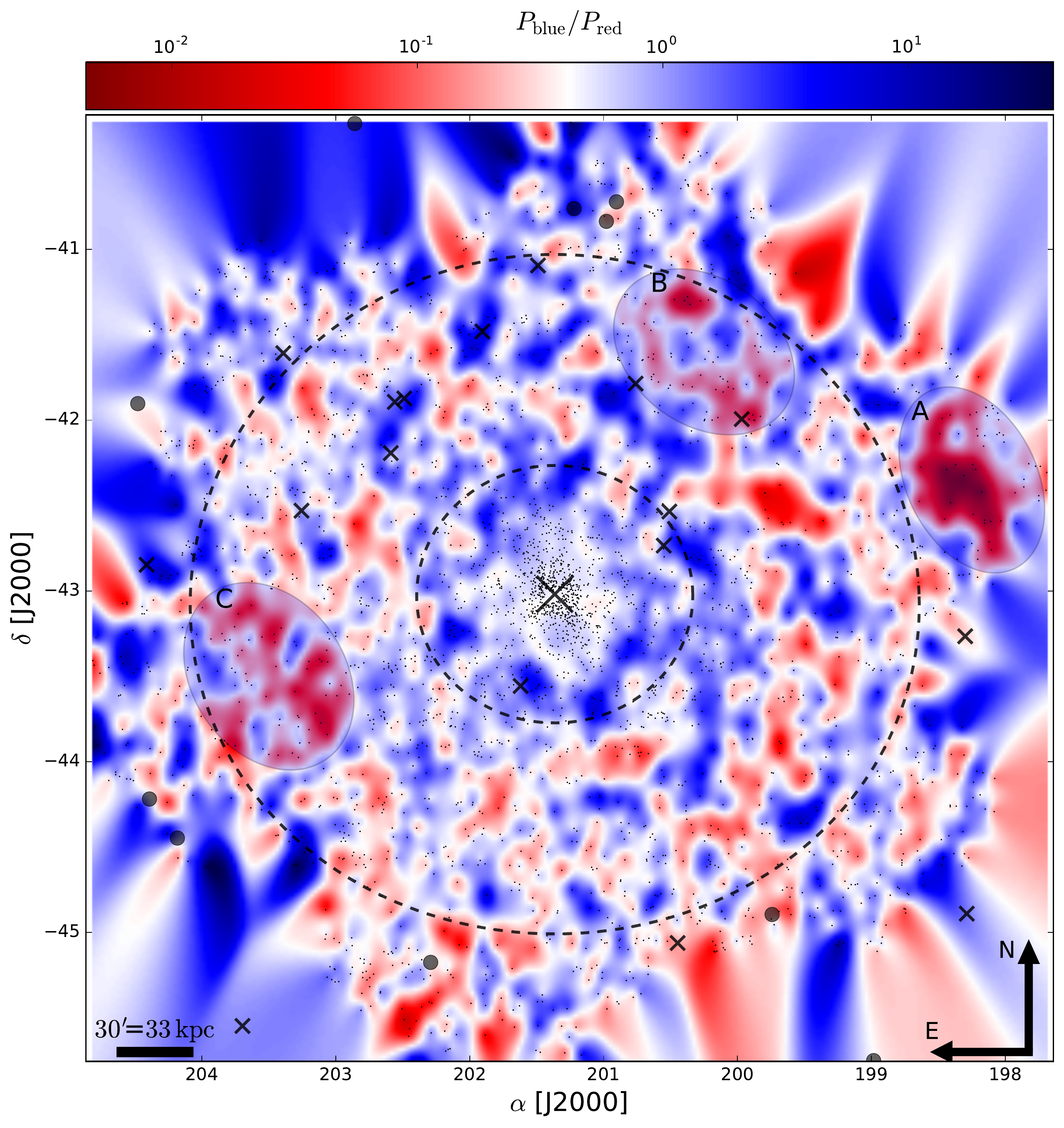}
\caption{The projected spatial probability density of blue GCs relative to red. Relative probability densities are highlighted by the darker red and blue regions, with white areas indicating equal contributions of blue and red GCs. Small points show locations of the GC candidates, while confirmed and candidate dwarf galaxies are indicated by small X's and circles, respectively. Dashed ellipses are drawn at $R_{\rm gc}=50\arcmin$ and $120\arcmin$, while lightly shaded ellipses highlight regions in \cena's outer halo with unexpectedly high densities of red GC candidates..}
\label{fig:btr_dense}
\end{figure*}

Figs.\,\ref{fig:inout_frac} and \ref{fig:btr_dense} further probe the spatial distributions of the ``inner'' and ``outer'' samples of blue and red GCs. The upper panels of Fig.\,\ref{fig:inout_frac} show two-component GMM fits to the ``inner'' GCs (left), and the ``outer'' candidates (right). Both samples show larger than unity \btrfrac, marginally for the outer candidates, and more dramatically for the inner. Meanwhile, the lower panel of Fig.\,\ref{fig:inout_frac} shows the $\Phi$ dependence of \btrfrac. The total (black relation), inner (green), and outer (brown) candidate samples are binned by $\Delta\Phi=36^\circ$, and \btrfrac, shown along the ordinate, is calculated in each bin. The outer sample shows on average lower \btrfrac, especially in the range $0^\circ\lesssim \Phi\lesssim 225^\circ$, and particularly at $\Phi\approx325^\circ-25^\circ$, $\sim95^\circ$, and $\sim200^\circ$.

Complementary to Fig.\,\ref{fig:inout_frac}, Fig.\,\ref{fig:btr_dense} shows the ratio of exponential-KDEs applied as in Fig.\,\ref{fig:gcdense}, but to each of the blue and red subsamples. The colour shading represents areas where the GCs of each colour dominate the surface number density distribution. Of particular note are the high \btrfrac\ features mentioned above. While it can be seen that \btrfrac\ is indeed generally higher inwards of $R_{\rm gc}\!\approx\!50\arcmin$, the blue GCs tend toward the outskirts of the region, with a particularly high concentration in the $130^\circ\lesssim\!\Phi\!\lesssim 250^\circ$ wedge S of \cena, but overall the central regions show a stronger representation by the red GC sample.

As seen previously, the region within $R_{\rm gc}\approx15\arcmin$ is represented well by each colour, and is surrounded by clumpy blue over densities corresponding to the \btrfrac\ spikes seen in Figs.\,\ref{fig:surfazi} and \ref{fig:inout_frac}. At larger $R_{\rm gc}$, both populations cluster at the scales indicated by Fig.\,\ref{fig:2pt}, with half of the dwarf galaxies showing some indications of projected associations with regions of blue over-densities.

%%%%%%%%%%%%%%%%%%%%%%%%%%%%%%%%%%%%%%%%%%%%%%%%%%
%%%%%%%%%%%%%%%%%%%%%%%%%%%%%%%%%%%%%%%%%%%%%%%%%%
%%%%%%%%%%%%%%%%%%%%%%%%%%%%%%%%%%%%%%%%%%%%%%%%%%

\subsection{Formation Scenarios for \cena's Globular Cluster System}
\label{sec:gctotal}
In the following, we explore the implications that the respective ``inner'' and ``outer'' GC candidate populations have for the past and present dwarf galaxy populations of \cena. We note that, while the apparent break in the power-law fits shown in Fig.\,\ref{fig:surfazi} might imply a transition to the Centaurus\,A intra-group medium, this may not reflect the true nature of these objects. For example, observational and theoretical evidence suggests that at least some giant galaxies reside within dual-halos, including a diffuse metal-poor stellar component extending dozens of kpc from their hosts that arise from either two-component star formation/chemical enrichment mechanisms \citep[e.g.][]{rej05,har07,rej11,lee16}, and/or from the accretion of multiple satellites throughout a galaxy's formation history \citep[e.g.][]{bul05,aba06,joh08,coo10,dea13,par13,iba14}. With this in mind we move forward to consider separately the inner, intrinsic, GC population of \cena, followed by the outer group that likely reflects GCs accreted onto the host's extreme outer halo and/or a transition to the Centaurus\,A intra-group GC population.

\subsubsection{The Inner GC Population}
\label{sec:innergcs}
If the inner GCs at $R_{\rm gc}\!<\!50\arcmin$, corresponding to $\sim\!55$\,kpc or $\sim20$ percent of \cena's virial radius, are intrinsic to \cena, they have strong implications for the assembly of the host. The median $(g'-z')_0=1.27\pm0.01$\,mag colour of the inner red GCs is $0.03$\,mag bluer than the $1.30\pm0.02$\,mag median $(g'-z')_0$ colour expected of a giant galaxy sharing \cena's $M_B\!=\!-20.1$\,mag luminosity in a cluster environment, and more consistent with a gE galaxy with $M_B$ in the range $-19.25$ to $-19.75$\,mag and $(g'-z')_0=(1.27-1.29)\pm0.02$\,mag \citep{pen06}. A merger of such galaxies would produce a combined luminosity of $M_B\approx-20.0$ to $-20.5$\,mag. This range brackets \cena's luminosity, consistent with the notion that \cena\ experienced a roughly equal mass merger in the recent past \citep[e.g.][]{baa54,gra79,inn79,tub80,mal83,hes86,qui93,min96,sti04}. Still, the prevalence of blue GCs in the halo of \cena\ out to $R_{\rm gc}\approx50\arcmin$ requires an explanation even if they were brought in during the merger event.

The new and H12 candidates within $50\arcmin$ together with the confirmed GCs implies an intrinsic total population of at least 1\,066. We can then calculate the specific frequency as defined by \citep{har81},
\begin{equation}
\label{eq:sfreq}
S_N = N_{GC}\cdot 10^{0.4\left(M_V+15\right)}=\left(8.51\times10^7\right)\frac{N_{GC}}{L_V/L_{V,\odot}}
\end{equation}
Adopting a de-reddened $V$-band luminosity of $M_V=-21.4$ with Eq.\,\ref{eq:sfreq} gives $S_N\approx2.9$, a value not unreasonable for a giant E/S0 galaxy \citep{har81,har13}. We note that the distribution of the red GC component clusters more closely to \cena\ than the blue GCs, which extend well out into the $R_{\rm gc}<50\arcmin$ halo with a shallower $\Sigma_N(R_{\rm gc})$ profile. This is consistent with other high $S_N$ galaxies, in that they tend to have higher \btrfrac, with metal-poor (blue) GCs preferentially following more radially extended, shallow distributions and extending further out than the host's underlying starlight \citep{for97b}.

\subsubsection{Hierarchical Assembly of the GC System}
While the consistency with previous works is encouraging, the relatively high \btrfrac\ calls for an explanation of its origins. The number of blue GCs that need to be accounted for is simply,
\begin{equation}
N_{\rm blue}=\frac{\xi_{\rm b/r}N_{GC,t}}{1+\xi_{\rm b/r}}
\end{equation}
If minor mergers with \cena's giant progenitors represent the sole sources of the blue GCs, then the number of former dwarfs, $N_{\rm dw}$, of a luminosity, $L_{V, {\rm dw}}$, needed to build up such population is approximated by the number of GCs that each galaxy can be expected to contribute, $N_{\rm GC,dw}$. From the definition of $S_N$, then trivially,
\begin{equation}
N_{\rm dw}=\frac{N_{\rm blue}}{N_{\rm GC,dw}}=\frac{\xi_{\rm b/r}N_{\rm GC,t}}{1+\xi_{\rm b/r}}\cdot\frac{8.51\times10^7}{S_{N,{\rm dw}}L_{V,{\rm dw}}}
\end{equation}
which can be simplified given \btrfrac$=1.33$ (c.f.\ Table\,\ref{tbl:gccolours}) for the 1\,066 inner \cena\ GCs to,
\begin{equation}
N_{\rm dw}=\frac{5.18\times10^{10}}{S_{N,{\rm dw}}L_{V,{\rm dw}}}
\end{equation}

Taking approximate $S_{N, {\rm dw}}$ for a dwarf at a given $L_{V,{\rm dw}}$ from the rich compilation of $S_N$ derived for dwarf galaxies in the local universe by \cite{geo10}, Table\,\ref{tbl:dwarfs} shows the approximate numbers of dwarfs of a given $L_{V,{\rm dw}}$ that would be required to fully account for the inner blue GCs detected. Also listed are the magnitudes of the combined dwarf progenitors, and the fraction of \cena's current luminosity that such a reservoir of dwarfs would contribute. While the numbers are rough approximations, it is illustrative that a reservoir of $>250$ $10^{6-7}\,L_{V,\odot}$-class dwarfs is sufficient to provide the blue GCs, but insufficient to build up a significant fraction of \cena's current luminosity. Meanwhile, $\sim100$ $10^{8}\,L_{V,\odot}$-class dwarfs are required to provide such blue GC population, and could have contributed up to $\sim33$ percent of \cena's light. We note that if a smaller number of $10^9\,L_{V,\odot}$-class dwarfs is responsible, this would be sufficient to build up \cena's total luminosity, but would not provide a means of explaining the origins of the red GC population. This last fact suggests then, that at least dozens of minor mergers of $10^{6-8}\,L_{V,\odot}$ dwarf galaxies with giant galaxy progenitors are required to explain \cena's current state. 

\begin{table}
	\centering
	\caption{Estimates of the dwarf galaxy reservoirs required to provide the blue GC population of \cena. Col.\,(1) lists the example luminosity classes of the purported dwarfs, followed by the approximate $S_N$ in col.\,(2). Col.\,(3) shows the number of dwarfs required to provide the observed blue GCs, while the last two columns list approximations of the combined luminosities of each dwarf reservoir and the fractions of \cena's current light that they would represent.}
	\label{tbl:dwarfs}
	\vspace{1mm}
	\begin{tabular}{l*{2}{S[table-format=3.0]}*{1}{S[table-format=-2.2]}r}
		\hline\hline
		$L_{V,{\rm dw}}$ 	& 	$S_{N,{\rm dw}}$	&	$N_{\rm dw}$	&	$M_{V, {\rm dw}}$	&	$f_{L_{V}}$	\\
		($L_\odot$)	&				&				&	${\rm (mag)}$		&			\vspace{1mm}\\
		\hline\\
		\multicolumn{5}{c}{$R_{\rm gc}<50\arcmin$ ({\it inner GC system})} \vspace{1mm}\\ 
		$5\times10^6$		&	39	&	266		&	-18.0		&	0.04	\\
		$1\times10^7$		&	21	&	247		&	-18.7		&	0.08	\\
		$1\times10^8$		&	5	&	104		&	-20.2		&	0.33	\\
		$1\times10^9$		&	2	&	26		&	-21.2		&	0.84	\\
		\hline\\
		\multicolumn{5}{c}{$R_{\rm gc}\geq50\arcmin$ ({\it outer GC system})}\vspace{1mm}\\
		$5\times10^6$		&	39	&	468		&	-18.6		&	0.08	\\
		$1\times10^7$		&	21	&	435		&	-19.3		&	0.14	\\
		$1\times10^8$		&	5	&	183		&	-20.8		&	0.59	\\
		$1\times10^9$		&	2	&	46		&	-21.8		&	1.47	\\
		\hline\hline
	\end{tabular}
\end{table}

\begin{table*}
	\centering
	\caption{Results of dwarf galaxy population modelling for the inner ($R_{\rm gc}<50\arcmin$) and outer ($R_{\rm gc}\geq50\arcmin$) regions around \cena. Col.\,(1) lists the assumed Schechter function slopes, followed by the required dwarf galaxy populations in luminosity bins increasing from cols.\,(2)--(5). The total numbers of dwarf galaxies are listed in col.\,(6), followed by their combined luminosities and fractions of \cena's current day light in cols.\,(7)--(8).}
	\label{tbl:dwarfmix}
	\vspace{1mm}
	\begin{tabular}{*{1}{S[table-format=-1.2]}*{5}{S[table-format=3.0]}*{1}{S[table-format=-2.2]}r}
		\hline\hline
		$\alpha$ 	& 	$N_{10^{6-7}\,L_{V,\odot}}$	&	$N_{10^{7-8}\,L_{V,\odot}}$	&	$N_{10^{8-9}\,L_{V,\odot}}$	&	$N_{10^{9-9.3}\,L_{V,\odot}}$	&	$N_{\rm dw}$	&	$M_{V, {\rm dw}}$		&	$f_{L_{V}}$	\\
					&								&								&								&				&				&	${\rm (mag)}$			&				\\
		\hline\\
		\multicolumn{8}{c}{$R_{\rm gc}<50\arcmin$ ({\it inner GC system})} \vspace{1mm}\\ 
		0.00		&	0	&	0	&	9	&	11	&	21	&	-21.06	&	0.73	\\
		-0.25		&	0	&	2	&	12	&	10	&	25	&	-21.04	&	0.72	\\
		-0.50		&	1	&	5	&	15	&	9	&	31	&	-21.00	&	0.69	\\
		-0.75		&	6	&	10	&	18	&	8	&	42	&	-20.93	&	0.65	\\
		-1.00		&	21	&	20	&	20	&	6	&	68	&	-20.79	&	0.57	\\
		-1.25		&	63	&	33	&	18	&	3	&	119	&	-20.49	&	0.44	\\
		-1.50		&	136	&	40	&	12	&	1	&	191	&	-19.91	&	0.26	\\
		-1.75		&	209	&	34	&	6	&	0	&	249	&	-19.12	&	0.13	\\
		-2.00		&	255	&	23	&	2	&	0	&	280	&	-18.35	&	0.06	\\
		\hline\\
		\multicolumn{8}{c}{$R_{\rm gc}\geq50\arcmin$ ({\it outer GC system})}\vspace{1mm}\\
		0.00		&	0	&	1	&	17	&	19	&	38	&	-21.66	&	1.27	\\
		-0.25		&	0	&	3	&	21	&	18	&	43	&	-21.64	&	1.25	\\
		-0.50		&	2	&	8	&	26	&	16	&	54	&	-21.61	&	1.21	\\
		-0.75		&	10	&	18	&	32	&	13	&	75	&	-21.54	&	1.14	\\
		-1.00		&	37	&	36	&	35	&	10	&	120	&	-21.40	&	1.00	\\
		-1.25		&	111	&	59	&	32	&	6	&	210	&	-21.10	&	0.76	\\
		-1.50		&	240	&	70	&	22	&	2	&	336	&	-20.54	&	0.46	\\
		-1.75		&	368	&	60	&	10	&	0	&	440	&	-19.74	&	0.22	\\
		-2.00		&	450	&	40	&	3	&	0	&	494	&	-18.97	&	0.11	\\
		\hline\hline
	\end{tabular}
\end{table*}

We further explore the purported dwarf galaxy population needed to give rise to the blue GC candidates by modelling the galaxy luminosity function (LF) assuming a Schechter function of the form \citep{sch76},
\begin{equation}
\Phi(L,\alpha)dL=\phi^*\left(\frac{L}{L^*}\right)^\alpha\cdot e^{-\left(L/L^*\right)}dL\\
\end{equation}
or in terms of magnitudes,
\begin{equation}
\label{eq:dwarfmix}
\begin{split}
\Phi(M,\alpha)dM=0.4\ln{(10)}\times\hspace{4cm}\\
\hspace{1cm}\times\phi^*\left[10^{0.4\left(M^*-M\right)}\right]^{\left(\alpha+1\right)}\cdot e^{-10^{0.4\left(M^*-M\right)}}dM 
\end{split}
\end{equation}
where $M^*$ is the characteristic galaxy magnitude, which following \cite{smi09} and \cite{fer16} we adopt as $M_V^*\!=\!-20.84$\,mag. While the $\phi^*$ normalization is somewhat dependent on the filter and magnitude range of a given dataset, we assume a typical value of $\phi^*=1.2\times10^{-2}\,h^3\,{\rm Mpc}^3$ \citep{sch76,smi09} with $h=68\,{\rm (km/s)}\,{\rm Mpc}^{-1}$ \citep{pla14}. We employ this distribution for $-18.42\leq M_V\leq -10.17\,{\rm mag}$ (corresponding to $2\times10^{9}\geq L_{V,\odot}\geq10^6$). Testing $\alpha$ in the range $(-2.0,0.0)$, we randomly draw a dwarf galaxy from the population and estimate the number of blue GCs that it could contribute to the total samples inside and outside of $50\arcmin$ of \cena. We repeat this process until $N_{\rm GC,blue}$ are contributed, and record the full dwarf sample. We iterate this procedure 1\,000 times for a given $\alpha$ and list the mean numbers of $10^{6-7}$, $10^{7-8}$, $10^{8-9}$, and $10^{9-9.3}\,L_{V,\odot}$-class dwarfs\footnote{We choose $10^{9.3}\,M_\odot$ as the upper sampling limit as it marks the highest luminosity of known dwarfs around \cena.} required to build the blue GC population in Table\,\ref{tbl:dwarfmix} at varying $\alpha$. We also show the total numbers of dwarfs predicted, alongside their combined magnitudes and fractions of \cena's current day luminosity.

Based on the results for the inner GC candidates listed in Table\,\ref{tbl:dwarfmix}, we find that a top-heavy LF ($\alpha\simeq0.0$) is disfavoured. Such a form predicts that as many as 11 dwarfs of $\gtrsim10^9\,L_{V,\odot}$ may have contributed to the observed blue GC candidates, but with little to no contribution by lower mass dwarfs. More importantly, the combined luminosity of such a dwarf population represents $\gtrsim73$ percent of \cena's spheroid light, but fails to provide an avenue for the creation of the red GC population, assuming them to have been formed during the monolithic collapse of \cena's giant progenitors. Distributions more top-heavy than $\alpha>-1.0$ tell a similar story, with $f_{L_V}\gtrsim0.65$ in all cases. For $-1.50\lesssim\alpha\lesssim-1.0$, we find that the increased fractions of blue GCs provided by lower-mass dwarfs sufficiently explain the GC population, while providing a fraction of \cena's current day light on the order of 25--60 percent. At more bottom-heavy dwarf LFs ($\alpha\lesssim-1.50$), the number of bright dwarfs would be expected to be zero, with $\gtrsim200$ $10^{6-7}\,L_{V,\odot}$-class dwarfs being required to provide the blue GCs, while providing only a small fraction of \cena's current spheroid light.

Altogether, these results suggest that a faint-end LF slope of $-1.25\lesssim\alpha\lesssim-1.50$ is consistent with the build-up of \cena's intrinsic population of blue GCs. This outcome is in good agreement with the recent work on the inner $\sim300$\,kpc of the Virgo galaxy cluster by \cite{fer16}, who found $\alpha=-1.33\pm0.02$ to best represent the LF in the dwarf regime, and is only marginally higher than their $\alpha=-1.21\pm0.05$ found for the Local Group dwarf galaxy population. Interestingly, \citeauthor{fer16} find that if the Virgo UCDs are assumed to be the remnants of nucleated dwarf galaxies stripped of their outer stellar halos (and thus GCs), they find a much steeper slope of $-1.60\pm0.06$, which appears inconsistent with our results. With that said, our GC selection procedure intentionally filtered out objects encroaching upon UCD luminosities, and so a future dedicated search for UCDs in the outer halo of \cena\ may result in a promising reservoir of additional blue GC progenitors.

\subsubsection{The Extreme Halo of \cena\ and its GC system}
\label{sec:outergcs}
Most of the dwarfs contributing to the inner population of GCs are likely to have merged with \cena\ and/or its giant progenitors in the past, but the same cannot necessarily be assumed for the extreme outer halo of \cena. In this way, carrying out the exercise above with the outer population serves as a rough prediction on the population of dwarfs that have either already been disrupted, or may still be present in the extreme halo. The bottom halves of Tables\,\ref{tbl:dwarfs} and \ref{tbl:dwarfmix} list these results, which predict an even richer dwarf population than the inner halo. If the LF of dwarfs in the extreme halo of \cena\ follows that within $R_{\rm gc}<50\arcmin$, this result requires that $>100$ $10^{6-7}\,L_{V,\odot}$ and dozens of $10^{7-9}\,L_{V,\odot}$-class dwarfs with projected $R_{\rm gc}\gtrsim50\arcmin$, many of which might have already been disrupted in \cena's tidal field, while their GC populations survived. The most massive of these dwarfs must still be or have until the recent past been present. 

The Centaurus\,A/M83 galaxy complex has been shown recently to be potentially rich in low-mass dwarf galaxies, with 57 promising candidates reported based on wide field DECam imaging \citep[][]{mul15,mul17} and 15 confirmed dwarfs within the region studied in this work \citep[][]{van00,kar07,crn14,crn15}; however, very little wide-field imaging has been done at sufficient depths to robustly detect dwarfs of $L_{V}\lesssim10^6\,L_{V,\odot}$. Excellent work was recently done to detect several new dwarfs \citep{crn15} via resolved stellar over densities, but the faintest of this sample barely reaches $L_{V}\approx10^6\,L_{V,\odot}$. Any dwarfs of similar luminosities, but more diffuse morphologies and/or projected along less-fortunate axes (assuming triaxial structures), would therefore still remain undetected, as would more massive satellites with yet more diffuse stellar distributions. With that said, \citeauthor{crn15} detected clear signals of loops and streams that indicate that dwarfs are still being actively disrupted to the present day out to at least $R_{\rm gc}\approx120\arcmin$ ($\sim132$\,kpc), thus it is likely that past dwarf galaxy interactions may be behind the population of blue GCs at such distance from \cena. Such interactions have recently been shown to produce preferential stripping of GC systems \citep{smi13, smi15}.

Even so, given the emerging evidence for at least one, and possibly two, planes of dwarf satellites around \cena\ \citep{tul15,mul16}, one might expect that if the outer GCs arise from accreted dwarf GC systems, then they should reflect this origin by assuming a coherent structure aligned with the plane associated with \cena\ itself (see \citeauthor{mul16}, their Fig.\,1). A planar alignment of outer GCs is clearly not seen in Fig.\,\ref{fig:gcdense} nor, in particular, the lower panel of Fig.\,\ref{fig:surfazi}, where the azimuthal distribution indicates large-scale near homogeneity. While it is tempting to attribute this as evidence against dwarf galaxy origins, the \cena-centric spatial sampling of only $\sim120$\,kpc precludes this interpretation, considering the Mpc-scale distribution of the satellite planes.

An interesting feature of Fig.\,\ref{fig:btr_dense} is the existence of several $\gtrsim\!30\arcmin$-scale structures dominated by red GCs, highlighted by shaded ellipses and labelled `A', `B', and `C'. The most significant (`A') lies $>\!120\arcmin$ to the NW of \cena\ and the existence of these GCs, along with those that make up the two red over-densities at $(90\,{\rm kpc},315^\circ)$ and $(R_{\rm gc},\Phi)\!\approx\!(90\,{\rm kpc},100^\circ)$ (labelled `B', and `C', respectively) is puzzling and merits follow-up studies. These GC overdensities may have origins in giant background galaxies or distant galaxy clusters, but visual inspection of the images shows no obvious evidence for possible background hosts. 

Evidence that the kinematics of the inner halo GCs shares similarities with the overall kinematics of the Centaurus A group suggests that the outer halo of the group is dynamically connected to the rest of the group \citep{woo06}. In this scenario, the notion that \cena\ has primarily been built up by minor mergers, with only a few major mergers contributing seems to be consistent with the present findings. While the results of the previous section support a recent merger of two equal-mass giants, it does not rule out a more ancient merger whose evidence is no longer detectable. In this case, these clumps could represent the last coherent structures resulting from these violent events. On the other hand, the recent discovery of ``ultra-diffuse'' galaxies (UDGs) in large galaxy clusters \citep{dok15,kod15,mih15,mun15,mar16} have been proposed to represent failed $L^*$ galaxies with deep potentials relative to their stellar masses. Little is known about their respective GC systems, but recent work has indicated that they may have high $S_N\gtrsim30$ with as many as dozens of GCs \citep{bea16,pen16,dok16}. These works find that UDGs are likely to host primarily blue GC populations more similar to dwarfs than giants, although with only one UDG GC system studied so far, few general conclusions can be drawn. In any case, the large GC substructures seen in Figs.\,\ref{fig:gcdense} and \ref{fig:btr_dense} could represent very interesting targets for follow-up deep imaging campaigns that would further build upon this work.

%%%%%%%%%%%%%%%%%%%%%%%%%%%%%%%%%
%%%%%%%%%%%%%%%%%%%%%%%%%%%%%%%%%
%%%%%%%%%%%%%%%%%%%%%%%%%%%%%%%%%

\section{Conclusions}
\label{sec:conc}
In this work, new wide-field CTIO/DECam imaging in the optical $u'g'r'i'z'$ filters of the central $\sim21\,{\rm deg}^2$ of the Centaurus A galaxy group is analyzed. Two-colour diagnostic diagrams are combined with source morphologies to construct a near-complete catalogue of GC candidates as far out as $\sim\!140$\,kpc from the centrally dominant galaxy \cena. We find a total of 2\,676 GC candidates, of which 2\,404 are newly identified, and provide new measurements of many previously radial-velocity confirmed GCs.

We use Gaussian mixture models to classify the GC candidates as either blue or red, and find the well established bimodal distribution in GC colours for giant galaxies is well defined for \cena's GC system. The GC system as a whole shows a larger number ratio of blue GCs with respect to red, i.e.\ \btrfrac$=1.16$. Evidence is presented for distinct populations inside and outside of $R_{\rm gc}\approx50\arcmin$ and we suggest that the inner population is likely intrinsic to \cena\ itself, while the outer population may begin to sample those GCs associated with the intra-group environment, or at least have been deposited there by accreted satellites. The inner sample shows a slightly higher \btrfrac$=1.33$, compared to $1.08$ for the outer, noting that the outer sample is likely to be more strongly affected by a bias against blue GCs due to our selection process. No evidence for a trimodal colour distribution \citep[see][]{woo10b} is found in our data, likely due to our vastly larger sample size.

We calculate $\Sigma_N(R_{\rm gc})\propto R_{\rm gc}^\Gamma$ for \cena's GC system and find that the inner population is consistent with other gE galaxies both in colour-dependent slopes of $-1.8\lesssim\Gamma\lesssim-1.40$ \citep[e.g.][]{puz04}, and that the red GCs are more centrally concentrated than the blue, which extend into \cena's outer halo. This trend reverses outside of $\sim50\arcmin$, as the blue GCs show a slightly steeper power-law relation. Overall, the spatial distribution of \cena's GC system is not uniform. A two-point angular correlation function analysis provides evidence for clustering on all scales below $\sim20$\,kpc for the inner population, with red GCs showing stronger clustering toward smaller scales. Mild evidence for a reversal of this trend is seen for the outer population, as the blue population shows slightly stronger evidence for small-scale clustering, consistent with the notion that they may be hinting at an as-yet unknown population of low surface-brightness dwarf galaxy hosts. Alternatively, they might be the last remaining coherent structures from previously disrupted dwarf satellites, as observed in the Local Group \citep{mac14}. Both red and blue outer GC samples show a much more shallow decline of $w(\theta)$, suggestive of clustering at larger scales than found in the inner population. Finally, we find mild evidence for a coherent over density, or stream, of GCs outside of $R_{\rm gc}\approx50\arcmin$, which will be statistically quantified in a future work.

The median $(g'-z')_0\approx0.90$\,mag colour of the blue component is consistent with what is expected for a gE galaxy of similar luminosity in the Virgo cluster \citep{pen06}, while the $\sim\!1.27$\,mag colour of the red component is consistent with a build up from two giant galaxies each on the order of $M_V\approx-19.2$ to $-19.8$\,mag. Based on the observed numbers of blue GCs within $\sim50\arcmin$ of \cena, if the assumption is made that they all have origins in dwarf haloes, this would require dozens of minor-mergers with $10^6-10^8\,L_{V,\odot}$-class dwarfs during the assembly of \cena\ and its giant progenitors. Likewise, if the outer population of blue GCs is to be explained by dwarf halo hosts, then yet more low-mass dwarfs are either possibly lying undetected, or have already been disrupted within $\sim140$\,kpc of \cena.

The unexpected presence of large numbers of both red and blue GCs in the extreme halo of \cena\ provides rich opportunities for follow up studies, which will be conducted in future contributions. Our SCABS imaging is undergoing a careful background subtraction and future work will characterize the as-yet relatively unknown dwarf population in the Centaurus A galaxy group. Additionally, the inclusion of near-infrared imaging will refine our new GC candidate catalogue further, with very low contamination by foreground stars and background galaxies \citep{mun14}. Nonetheless, with the complete census of true GCs soon to be in-hand, secure spectroscopic follow-up targets will be paramount to unveiling the global velocity map of Centaurus A and its environment, and place unprecedented constraints on the mass assembly history of this iconic galaxy.

\section*{Acknowledgements}
We extend our gratitude to Marina Rejkuba for providing extensive and constructive comments that significantly improved the manuscript. We also wish to warmly thank Simon \'Angel, and Yasna Ordenes-Brice\~no for fruitful discussions, and especially Eric Peng for additionally providing us with catalogues of new confirmed foreground stars and GCs prior to publication. M.A.T.\ acknowledges the financial support through an excellence grant from the ``Vicerrector\'ia de Investigaci\'on" and the Institute of Astrophysics Graduate School Fund at Pontificia Universidad Cat\'olica de Chile and the European Southern Observatory Graduate Student Fellowship program. T.H.P. acknowledges support by a FONDECYT Regular Project Grant (No.~1161817) and the BASAL Center for Astrophysics and Associated Technologies (PFB-06).~H.Z. was supported in part by FONDECYT Postdoctoral Fellowship Grant (No. 3160538).~P.E.~acknowledges support from of a FONDECYT Postdoctoral Fellowship Grant (No. 3130485) and the CHINA-CONICYT Fellowship Project (No. CAS150023).~M.S.B. was supported in part by a FONDECYT Postdoctoral Fellowship Grant (No. 3130549).\\
This research has made use of the NASA Astrophysics Data System Bibliographic Services, the NASA Extragalactic Database, and the SIMBAD database, operated at CDS, Strasbourg, France \citep{wen00}.. Software used in the analysis includes the {\sc Python/NumPy} v.1.11.0 and {\sc Python/Scipy} v0.17.0 \citep[][\url{http://www.scipy.org/}]{jon01,van11}, {\sc Python/astropy} \citep[v1.1.1;][\url{http://www.astropy.org/}]{ast13}, {\sc Python/matplotlib} \citep[v1.5.1;][\url{http://matplotlib.org/}]{hun07}, {\sc Python/scikit-learn} \citep[v0.16.1;][\url{http://scikit-learn.org/stable/}]{ped12}, and {\sc Python/astroML} \citep[v0.3;][\url{http://www.astroml.org/}]{van12} packages.\\
This work is based on observations at Cerro Tololo Inter-American Observatory, National Optical Astronomy Observatory (CNTAC Prop. ID: 2014A-0610; PI: Matthew Taylor), which is operated by the Association of Universities for Research in Astronomy (AURA) under a cooperative agreement with the National Science Foundation.
This project used data obtained with the Dark Energy Camera (DECam), which was constructed by the Dark Energy Survey (DES) collaboration.

%%%%%%%%%%%%%%%%%%%%%%%%%%%%%%%%%%%%%%%%%%%%%%%%%%%
%
%%%%%%%%%%%%%%%%%%%%% REFERENCES %%%%%%%%%%%%%%%%%%
%
%% The best way to enter references is to use BibTeX:
%
%%\bibliographystyle{mnras}
%%\bibliography{example} % if your bibtex file is called example.bib
%
%
%% Alternatively you could enter them by hand, like this:
%% This method is tedious and prone to error if you have lots of references
\setlength\parskip{0.01\baselineskip}

%%%%%%%%%%%%%%%%%%%%%%%%%%%%%%%%%%%%%%%%%%%%%%%%%%

%%%%%%%%%%%%%%%%% APPENDICES %%%%%%%%%%%%%%%%%%%%%

%\appendix
%
%\section{Some extra material}
%
%If you want to present additional material which would interrupt the flow of the main paper,
%it can be placed in an Appendix which appears after the list of references.

%%%%%%%%%%%%%%%%%%%%%%%%%%%%%%%%%%%%%%%%%%%%%%%%%%

% Don't change these lines
\bsp	% typesetting comment
\label{lastpage}
\end{document}